\documentclass[useAMS,usedcolumn,usenatbib,usegraphicx]{mn2e}

\usepackage{epsfig,graphicx,natbib}
\usepackage{./references/mycite}
\usepackage{amssymb}
\usepackage{gensymb}
\usepackage{amsfonts}
\usepackage{amsmath}
\usepackage{color}
\usepackage{lscape,longtable}
 \usepackage{times}
 
\newcommand{\aap}{A\&A}

\newcommand{\apjl}{ApJL}
\newcommand{\apjs}{ApJS}

\begin{document}

\title[Gas Rich Starbursts in Luminous Reddened Quasars]{The Discovery of Gas Rich, Dusty Starbursts in Luminous Reddened Quasars at $z\sim2.5$ with ALMA}

\author[M. Banerji et al.]{ \parbox{\textwidth}
{Manda Banerji$^{1,2}$\thanks{E-mail: mbanerji@ast.cam.ac.uk}, 
C. L. Carilli$^{3,4}$,
G. Jones$^{3,5}$, 
J. Wagg$^{6}$,  
R. G. McMahon$^{1,2}$,
P. C. Hewett$^{1}$,
S. Alaghband-Zadeh$^{1}$, 
C. Feruglio$^{7}$ 
}
\\
$^{1}$Institute of Astronomy, University of Cambridge, Madingley Road, Cambridge CB3 0HA, UK\\
$^{2}$Kavli Institute of Cosmology Cambridge, Madingley Road, Cambridge CB3 0HA, UK\\
$^{3}$National Radio Astronomy Observatory, 1003 Lopezville Road, Socorro, NM 87801, USA\\
$^{4}$Battcock Centre for Experimental Astrophysics, Cavendish Laboratory, Cambridge CB3 0HE, UK \\
$^{5}$Physics Department, New Mexico Institute of Mining and Technology, Socorro, NM 87801, USA \\
$^{6}$SKA Organization, Lower Withington Macclesfield, Cheshire SK11 9DL, UK \\
$^{7}$Scuola Normale Superiore, Piazza dei Cavalieri 7, 56126, Pisa, Italy \\
}



\maketitle

\begin{abstract}
We present ALMA observations of cold dust and molecular gas in four high-luminosity, heavily reddened (A$_{\rm{V}} \sim 2.5-6$ mag) Type 1 quasars at $z\sim2.5$ with virial M$_{\rm{BH}} \sim 10^{10}$M$_\odot$, to test whether dusty, massive quasars represent the evolutionary link between submillimetre bright galaxies (SMGs) and unobscured quasars. All four quasars are detected in both the dust continuum and in the $^{12}$CO(3-2) line. The mean dust mass is 6$\times$10$^{8}$M$_\odot$ assuming a typical high redshift quasar spectral energy distribution (T=41K, $\beta$=1.95 or T=47K, $\beta$=1.6). The implied star formation rates are very high - $\gtrsim$1000 M$_\odot$ yr$^{-1}$ in all cases. Gas masses estimated from the CO line luminosities cover $\sim$1-5$\times10^{10}$($\alpha_{\rm{CO}} / 0.8$)M$_\odot$ and the gas depletion timescales are very short - $\sim5-20$Myr. A range of gas-to-dust ratios is observed in the sample. We resolve the molecular gas in one quasar - ULASJ2315$+$0143 ($z=2.561$) - which shows a strong velocity gradient over $\sim$20 kpc. The velocity field is consistent with a rotationally supported gas disk but other scenarios, e.g. mergers, cannot be ruled out at the current resolution of these data. In another quasar - ULASJ1234+0907 ($z=2.503$) - we detected molecular line emission from two millimetre bright galaxies within 200 kpc of the quasar, suggesting that this quasar resides in a significant over-density. The high detection rate of both cold dust and molecular gas in these sources, suggests that reddened quasars could correspond to an early phase in massive galaxy formation associated with large gas reservoirs and significant star formation. 

\end{abstract}




\section{Introduction}

Since the discovery of the correlation between the stellar bulge mass in galaxies and the mass of their supermassive black holes \citep{Magorrian:98, Kormendy:13}, it is now widely acknowledged that these central supermassive black holes play an important role in governing the formation and evolution of their host galaxies. In our current picture of galaxy formation, galaxies and their supermassive black-holes co-evolve. Molecular gas and the dusty interstellar medium (ISM) in galaxies, acts as the fuel for both star formation and black hole accretion. Therefore, studying the distribution of dust and gas in star forming galaxies with actively accreting supermassive black holes can help shed light on the exact physical processes that are driving co evolution. 

Both star formation and black hole accretion activity in the Universe peak at $z\sim2-3$ (e.g. \citealt{Madau:98, Richards:06}) and this represents the key epoch in the Universe's history at which to study the feeding and feedback processes in galaxies and quasars that are the main drivers of co-evolution. Over the last two decades, surveys starting with the SCUBA bolometer on the James Clerk Maxwell Telescope (JCMT) and, more recently, using the \textit{Herschel} space satellite, have led to the discovery of large populations of far infrared (FIR) and submillimetre bright starburst galaxies (SMGs) at the main epoch of galaxy formation (e.g. \citealt{Smail:97, Blain:99, Casey:14}). Molecular gas detections are now being assembled for many tens to hundreds of such galaxies (e.g. \citealt{Greve:05, Tacconi:06, Bothwell:13, Genzel:15}). Several studies of molecular gas in $z>2$ galaxies have also taken advantage of gravitational lensing to amplify the molecular line signal (e.g. \citealt{Coppin:07, Danielson:11, Aravena:16}). While clearly an advantage for probing intrinsically fainter galaxies, differential magnification between the different dust and gas components can make interpretations regarding their actual spatial distribution complicated in such lensed systems.

Quasars, on account of being among the brightest extragalactic sources in our Universe, were some of the first sources to be observed in molecular gas (e.g. \citealt{Solomon:05, Barvainis:97}). Using the compilations of extragalactic sources with CO detections in \citet{Carilli:13} and \citet{Heywood:13}, we find there are a total of 89 quasars that have been detected in CO all the way from the local Universe out to the highest redshifts of $z>6$. Many of the recent efforts have focussed on $z>4$ quasars (e.g. \citealt{Carilli:02, Walter:04, Wang:10}) and, as a consequence, the numbers of quasars with CO detections at the peak epoch of galaxy formation (1.5$<z<3.0$) is considerably smaller - $<$20 quasars. Many of the quasars that have been observed in molecular gas at these epochs are gravitationally lensed systems (e.g. \citealt{Riechers:11b, Sharon:16}). Excluding these lensed systems and focussing only on unlensed quasars where spatially mapping the gas and dust distributions is considerably easier, there are only 13 quasars at $1.5<z<3.0$ with CO detections \citep{Coppin:08, Simpson:12, Schumacher:12, Willott:07}. Detecting molecular gas in even a small number of unlensed quasars at these redshifts, to synchronously study the fuelling of star formation and black hole accretion at the epoch when both are at their peak, is therefore valuable. 

Most searches for molecular gas in quasars have started from samples of ultraviolet (UV) luminous quasars (e.g. \citealt{Engels:98, Hewett:95, Schneider:10}) that were already known to be FIR/millimetre bright \citep{Omont:03, Priddey:03, Stevens:05}. The studies by \citet{Priddey:03} and \citet{Omont:03} find that only 9 out of 53 UV luminous quasars at $z\sim2$ (i.e. 17 per cent) are detected at 850$\mu$m down to 3$\sigma$ flux density limits of $\sim$7-9 mJy and 9 out of 26 UV luminous quasars at $z\sim2$ (i.e. 35 per cent) are detected at 1.2 mm down to 3$\sigma$ flux density limits of $\sim$1.8-4 mJy. In the galaxy formation scenario first advocated by \citet{Sanders:88} and now commonly adopted in galaxy formation models (e.g. \citealt{Hopkins:08, Narayanan:10}), the most highly star forming galaxies at high redshift e.g. the SMGs, will eventually evolve into UV luminous quasars. The relatively small fraction of FIR/mm-bright quasars argues for a quick transition between the two populations and recent studies suggest that the transition phase could be as short lived as $\sim$1 Myr \citep{Simpson:12}. In such a picture of galaxy formation, transition populations of hybrid SMG-quasars should exist and a larger fraction of these transition quasars should be FIR bright, gas rich systems compared to the UV luminous quasars. Molecular gas observations of SMGs and UV luminous quasars have also revealed some differences between the two populations. While SMGs appear to have substantial amounts of low excitation gas \citep{Riechers:11a, Bothwell:13}, this does not seem to be the case in quasars \citep{Riechers:11b} which could therefore represent a later evolutionary stage when the extended gas reservoirs in the galaxy have been significantly depleted. Once again, if transition objects between these two populations do exist, their gas fractions might be expected to be intermediate between the SMGs and UV luminous quasars. 

Searches for these so called ``transition" galaxies have often focussed on obscured AGN populations with different selection methods picking out AGN and quasars with a wide variety of dust extinctions and luminosities (e.g. \citealt{Urrutia:08, Brusa:10, Banerji:12, Banerji:13, Banerji:15, Glikman:12, Eisenhardt:12, Tsai:15}). While in principle, any or all of these populations could represent the missing evolutionary link between star forming galaxies and optical quasars, in practice further multiwavelength observations are necessary to establish that these obscured quasars are indeed distinct in terms of their physical properties from matched control samples of unobscured quasars. Our searches for obscured quasars have been focussed at the highest luminosities and at the peak epoch of galaxy formation at $z=2-3$ with colour cuts deliberately chosen to isolate high luminosity quasars with the same levels of dust extinction as seen in SMGs at similar redshifts. To date, we have spectroscopically confirmed a new sample of almost 60 heavily reddened, luminous broad line quasars with a median A$_{\rm{V}} \sim$ 2.5 mags (c.f. A$_{\rm{V}} \sim 2.9\pm 0.5$ for SMGs; \citealt{Takata:06}). We have already found evidence that a higher fraction of our reddened quasars are actively star forming relative to optical quasars \citep{Banerji:14, Alaghband-Zadeh:16} consistent with these being transition objects seen as they are both rapidly forming stars and rapidly growing their black holes. Our reddened quasars are also among the most luminous quasars known at these epochs with recent results suggesting that their space density actually exceeds that of unobscured, UV luminous quasars at the highest luminosities \citep{Banerji:15}. The bolometric luminosities are $\sim$10$^{47}$ erg s$^{-1}$ and black hole masses are 10$^9$-10$^{10}$ M$_\odot$ inferred from the broad H$\alpha$ emission lines in the NIR spectra, consistent with the most luminous quasars found in optical surveys such as the Sloan Digital Sky Survey (SDSS).  

In order to determine whether the reddened broad line quasars are indeed a transition population between SMGs and optical quasars, direct comparison of their molecular gas and ISM properties to both SMGs and optical quasars is now necessary. Here we discuss the first ALMA observations of four heavily reddened quasars from \citet{Banerji:12} and \citet{Banerji:15} (B12 and B15 hereafter). Our new data is supplemented with two more detections of the $^{12}$CO(3-2) line in two quasars from our B12 parent sample \citep{Feruglio:14,Brusa:15} to give a sample of six heavily reddened quasars with molecular gas measurements. In Section \ref{sec:data} we present the ALMA observations of the four new quasars. Section \ref{sec:results} discusses both the dust continuum and $^{12}$CO(3-2) line properties of the quasars and presents physical properties of the quasar host galaxies including star formation rates, dust and molecular gas masses. In Section \ref{sec:discussion} we attempt to put our results in context with previous observations of molecular gas in both high redshift SMGs and optical quasars to determine whether the dust and gas properties of the reddened quasars are in fact intermediate between the two. Throughout this paper we assume a flat $\Lambda$CDM cosmology with h$_0$=0.7, $\Omega_M=0.3$, $\Omega_\Lambda=0.7$. 

\section{DATA}

\label{sec:data}

\subsection{Sample Selection \& ALMA Observations:}

In this work we make use of ALMA data obtained as part of the Cycle 3 project 2015.1.01247.S (PI:Banerji). The aim is to detect, for the first time, the molecular gas reservoirs in four heavily reddened quasars from B12 and B15 via the $^{12}$CO(3-2) emission line. We refer readers to B12, B15 and \citet{Banerji:13} for details of the parent sample of reddened (A$_{\rm{V}} \gtrsim 1.5$) broad emission line quasars from which our targets are drawn. Three of the four quasars selected for ALMA observations - ULASJ0123+1525, ULASJ1234+0907 and ULASJ2315+0143 - correspond to three of the reddest quasars in our sample with inferred dust extinctions towards the quasar continuum of A$_{\rm{V}}$=4.0, 6.0 and 3.4 mags respectively. Assuming the extinction is connected to star formation in the quasar host galaxy as discussed extensively in B12 and B15, these broad line quasars would be the best candidates for highly star forming gas rich hosts. The fourth quasar in this paper - VHSJ2101-5943 - represents a quasar with average extinction (A$_{\rm{V}}$=2.5 mags) in our sample. As such, it can be taken to be representative of the gas and dust properties of the larger sample of heavily reddened broad line quasars. We searched for radio counterparts to these quasars in the VLA-FIRST, NVSS and SUMSS radio catalogues and did not find any matches within 10$\arcsec$. The H$\alpha$ derived redshifts, AGN luminosities (at rest-frame 6$\mu$m) and black hole masses for these quasars can be seen in Table \ref{tab:properties}. 

We emphasise that apart from ULASJ1234+0907, which was detected in the \textit{Herschel} PACS and SPIRE bands \citep{Banerji:14}, none of the other three targets were known a-priori to be FIR or millimetre bright. This is in contrast to the targeting strategies for detecting molecular gas in UV luminous, unobscured quasars where only UV luminous quasars already known to be FIR bright have typically been observed in molecular gas. As discussed in B12 and \citet{Banerji:14}, it is unlikely that our reddened quasars are lensed. Optical and near infrared $J$ band images of our quasars demonstrate that the reddened quasars are extremely faint or completely invisible at these wavelengths and, given the depths of these imaging data, a lensing galaxy should have been visible out to $z>1$. 

\begin{table*}
\begin{center}
\caption{Summary of the properties - $K$ band magnitudes, redshifts, line-of-sight extinctions, AGN luminosities and black hole masses - for the four reddened quasars observed with ALMA. The properties have been derived from NIR photometry and spectroscopy and presented in B12 and B15.}
\label{tab:properties}
\begin{tabular}{lcccc}
& ULASJ0123$+$1525 & ULASJ1234$+$0907 & VHSJ2101$-$5943 & ULASJ2315$+$0143 \\
\hline
R.A. (J2000) & 01:23:12.52 & 12:34:27.52 & 21:01:19.46 & 23:15:56.23 \\
Dec (J2000) & $+$15:25:22.7 & $+$09:07:54.2 & $-$59:43:44.8 & $+$01:43:50.4 \\
K$_{\rm{AB}}$ & 18.59 & 18.05 & 16.68 & 18.38 \\
z$_{H\alpha}$ & 2.629 & 2.503 & 2.313 & 2.560 \\
A$_{\rm{V}}^\ast$ / mag & 4.0 & 6.0 & 2.5 & 3.4 \\
log$_{10}$(L$_{\rm{6\mu \rm{m}}}$ / erg s$^{-1}$) & 47.3 & 46.8 & 46.9 & 47.0 \\
log$_{10}$(M$_{\rm{BH}}$ / M$_\odot$) & 9.7 & 10.4 & 10.5 & 10.1 \\
\hline
\end{tabular}
\begin{small}
\begin{flushleft}
$^\ast$A$_{\rm{V}}$=R$_{\rm{V}}$$\times$$E(B-V)$=3.1$\times$$E(B-V)$
\end{flushleft}
\end{small}
\end{center}
\end{table*}

Observations were conducted between 2015 December and 2016 January and, as a detection experiment, made use of the most compact configuration available in Cycle 3, which corresponds to the lowest angular resolution ($\sim$2-3 arcsec). The number of useable antennae increased over the duration of the observations resulting in increased sensitivity for observations conducted at a later date. The correlator was configured to four dual polarization bands of 2 GHz (1.875 GHz effective) bandwidth each, providing a channel width of $\sim$15.6 MHz. The basebands were set up such that the $^{12}$CO(3-2) emission line is positioned in the first band with the other three bands providing a measurement of the dust continuum at observed wavelengths of $\sim$3 mm. The central frequencies of the four bands for each quasar, together with other details of the observations, can be found in Table \ref{tab:obs}. 

\begin{table*}
\begin{center}
\caption{Details of the ALMA observations for the four reddened quasars presented in this work.}
\label{tab:obs}
\begin{tabular}{lcccc}
& ULASJ0123$+$1525 & ULASJ1234$+$0907 & VHSJ2101$-$5943 & ULASJ2315$+$0143 \\
\hline
Date Completed & 2016-01-10 & 2015-12-26 & 2016-01-21 & 2016-01-07 \\
Exposure Time / sec & 3629 & 1058 & 4052 & 3145 \\
Number of Antennae & 46 & 34 & 48 & 47 \\
Beam Size (line) / arcsec & 3.0$\times$2.5 & 3.2$\times$2.6 & 3.6$\times$2.5 & 2.9$\times$2.2 \\
Observed Frequencies / GHz & 95.286, 94.286 & 98.714, 97.714 & 104.375, 103.267 & 97.134, 96.134 \\
& 107.286, 106.286 & 86.714, 85.714 & 92.375, 91.375 & 109.134, 108.134 \\
Channel RMS / mJy / beam$^{\ast}$  & 0.13 & 0.21 & 0.13 & 0.11 \\
Continuum RMS / $\mu$Jy / beam & 16 & 20 & 11 & 15 \\
Continuum Flux Density / $\mu$Jy / beam & 98$\pm$16 & 67$\pm$21 & 41$\pm$12 & 259$\pm$17 \\
z$_{\rm{CO}}$ & 2.6297$\pm$0.0012 & 2.5026$\pm$0.0012 & 2.3113$\pm$0.0034 & 2.5614$\pm$0.0028 \\
$^{12}$CO(3-2) Line Intensity / Jy km s$^{-1}$ & 1.40$\pm$0.08 & 0.97$\pm$0.18 & 0.46$\pm$0.03 & 0.91$\pm$0.05 \\
\hline 
\end{tabular}
\begin{small}
\begin{flushleft}
$^\ast$Median RMS over 15.6 MHz channels
\end{flushleft}
\end{small}
\end{center}
\end{table*}

\subsection{Data Reduction}

The ALMA data were calibrated using the ALMA pipeline in the Common Astronomy Software Applications package, \textsc{casa} (v4.5.1-4.5.2) by executing the appropriate ALMA calibration scripts corresponding to the release data of the observations. Time dependent amplitude and phase variations were calibrated using nearby quasars and radio galaxies. Flux calibrations made use of observations of Uranus and Neptune. The typical calibration uncertainties are of the order of 10-20 per cent. The beam sizes as well as the root mean square (RMS) sensitivity of both the $^{12}$CO(3-2) (median over 15.6 MHz channels) and the dust continuum observations can be seen in Table \ref{tab:obs}. Dust continuum images were produced for each quasar from the calibrated visibilities, by combining the line free channels from the three continuum spectral windows in multi-frequency synthesis mode using the CASA task \textit{clean} and a \textit{natural} weighting scheme to maximise the sensitivity. These same channels were also used to produce a UV plane model of the continuum emission, which was then subtracted from the first spectral window containing the line using the CASA task \textit{uvcontsub}. The continuum subtracted line visibilities were then imaged using \textit{clean}, once again employing a \textit{natural} weighting scheme in order to produce the final line cubes. Finally, both the continuum images and line cubes were corrected for the primary beam response.

The spectral profile in the data cubes is a convolution of the intrinsic spectrum with a spectral resolution function. By default, the ALMA pipeline calibration performs Hanning smoothing of the data cubes resulting in a spectral resolution of 2$\times$ the channel spacing. We have corrected for this effective reduction in resolution in all the spectral fits presented in this paper, although the effect is small. We have detected the dust continuum as well as the $^{12}$CO(3-2) line in all four quasars (Figs. \ref{fig:maps} and \ref{fig:spec}) and the dust continuum flux densities, CO line intensities and CO derived redshifts are presented in Table \ref{tab:obs}.     

\section{Results \& Analysis}

\label{sec:results}

\subsection{Dust Continuum: Dust Masses, FIR Luminosities \& Star Formation Rates}

\label{sec:dust}

The dust continuum detections at observed frame wavelengths of $\sim$2.9-3.5 mm (rest frame $\sim$800-900$\mu$m), suggest that there is a significant amount of cold dust in all four quasar host galaxies. The dust continuum emission is unresolved in all cases. The ALMA continuum detections probe the Rayleigh Jeans tail of the dust spectral energy distribution (SED) for these quasars, where the emission from the dust is expected to be optically thin. Under these assumptions dust masses can be calculated from the single photometric ALMA data point, $S_{\nu_{\rm{obs}}}$ after assuming a typical dust temperature, T$_{\rm{d}}$ and dust emissivity index, $\beta$ as follows:

\begin{equation}
M_{\rm{dust}}=\frac{{D_L}^2}{1+z} \times \frac{S_{\nu_{\rm{obs}}}}{\kappa_{\nu_{\rm{rest}}} B(\nu_{\rm{rest}},\rm{T}_{\rm{d}})}
\label{eq:mdust}
\end{equation}

\noindent where the mass absorption coefficient of the dust:

\begin{equation}
\kappa_{\nu_{\rm{rest}}}=\kappa_0 (\nu / \nu_0)^\beta
\label{eq:kappa}
\end{equation}

\noindent $D_L$ is the luminosity distance and $B(\nu_{\rm{rest}},\rm{T}_{\rm{d}})$ is the Planck function. We assume $\kappa_0$=0.045 m$^2$ kg$^{-1}$ at $\nu_0$=250 GHz \citep{Greve:12}. The uncertainty in the dust mass therefore depends primarily on the uncertainty in the dust temperature. In order to estimate the dust temperature, T$_{\rm{d}}$, and emissivity index, $\beta$, that is most appropriate for our quasars, we begin by using the fact that for one of our quasars, ULASJ1234+0907, we already have a complete sampling of the FIR spectral energy distribution (SED) from \textit{Herschel} observations \citep{Banerji:14}. The \textit{Herschel} data traces rest-frame wavelengths of $\sim$70$-$140$\mu$m in this quasar where the optically thin assumption tends to break down. We therefore fit a modified single temperature greybody to the \textit{Herschel} observations combined with the ALMA 3mm continuum detection assuming the emission is optically thick. The best-fit SED has parameters: T$_{\rm{d}}$=31$\pm$4K and $\beta$=2.5$\pm$0.3. 

\begin{figure}
\begin{center}
\vspace{-3cm}
\includegraphics[scale=0.4]{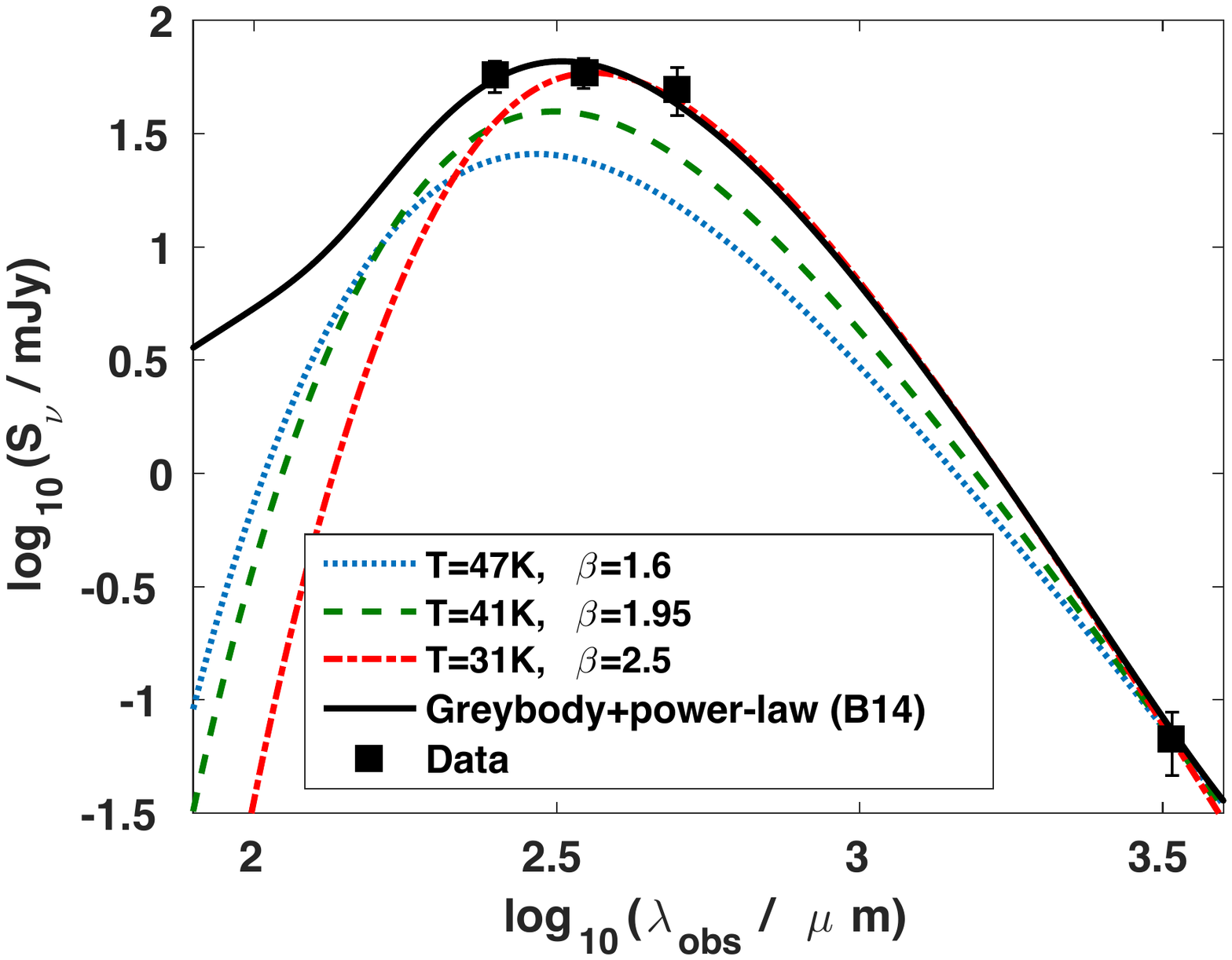}
\vspace{-3cm}
\caption{\textit{Herschel} SPIRE photometry at observed wavelengths of 250, 350 and 500$\mu$m together with the ALMA 3mm continuum photometry for ULASJ1234+0907 ($z=2.503$). The best fit power-law + single temperature greybody fit to these data points is also shown (see \citealt{Banerji:14}). We also plot three different single temperature, optically thick greybodies with T=31K, $\beta$=2.5 (best fit), T=41K, $\beta=1.95$ \citep{Priddey:01} and T=47K, $\beta=1.6$ \citep{Beelen:06, Wang:08}. In Section \ref{sec:1234} we show that the \textit{Herschel} fluxes are likely contaminated by other galaxies within the \textit{Herschel} beam. The latter two spectral energy distributions have therefore been used in the paper to calculate FIR luminosities, star formation rates and dust masses for all the reddened quasars in our sample.}
\label{fig:1234_sed}
\end{center}
\end{figure}

Previous investigations of the dust SEDs of high redshift quasars have found T$_{\rm{d}}$=47K and $\beta$=1.6 \citep{Beelen:06, Wang:08} or T$_{\rm{d}}$=41K and $\beta$=1.95 \citep{Priddey:01}. In Fig. \ref{fig:1234_sed}, we show all three assumptions for single temperature greybodies overlaid on the photometry for ULASJ1234$+$0907. As we shall see in Section \ref{sec:1234}, the \textit{Herschel} fluxes for ULASJ1234$+$0907 could be boosted by the presence of other millimetre bright galaxies within the \textit{Herschel} beam and the dust temperature and emissivity index derived from fitting the \textit{Herschel} + ALMA data may therefore not be accurate. Throughout this paper we will attempt to relate our measurements to the measured dust properties of other high redshift quasars. To be consistent with these previous works, from hereon we therefore choose to adopt the average of two commonly used assumptions for the dust SED of high redshift quasars: (i) T$_{\rm{d}}$=47K, $\beta$=1.6 and (ii) T$_{\rm{d}}$=41K, $\beta$=1.95. The dust masses in Table \ref{tab:derived} therefore correspond to dust temperatures of 41$-$47 K. 

Under the same assumptions for the dust SED, we can also calculate the FIR luminosities and star formation rates by simply scaling the optically thick single temperature greybodies to match the dust continuum detections at $\sim$3mm. The FIR luminosity here is defined between 40 and 300$\mu$m and we use the \citet{Kennicutt:12} relation to convert this FIR luminosity to a star formation rate:

\begin{equation}
\rm{SFR} / \rm{M}_\odot \rm{yr}^{-1} = 3.89\times10^{-44} \times \rm{L}_{\rm{FIR}} / \rm{erg s}^{-1}
\label{eq:sfr}
\end{equation}

\noindent Once again, these values can be found in Table \ref{tab:derived} and correspond to the average of the T$_{\rm{d}}$=47K, $\beta$=1.6 and T$_{\rm{d}}$=41K, $\beta$=1.95 greybodies. These assumptions should be kept in mind when interpreting the physical properties of the quasar host galaxies.

\begin{landscape}
\begin{table*}
\begin{center}
\caption{FIR luminosities, star formation rates, gas and dust masses for the four reddened quasars observed with ALMA and presented in this work, as well as the two reddened quasars from the same parent sample studied in \citet{Feruglio:14} and \citet{Brusa:15}. In the latter cases, the CO(1-0) luminosities and gas masses have been re-derived assuming the same CO excitation and CO-to-H$_2$ conversion factor as has been assumed for the quasars in this work.}
\label{tab:derived}
\begin{tabular}{lcccccc}
& ULASJ0123$+$1525$^\ast$ & ULASJ1234$+$0907$^\ast$ & VHSJ2101$-$5943$^\ast$ & ULASJ2315$+$0143$^\ast$ & ULASJ1539$+$0557$^+$ & ULASJ1002$+$0137$^\dagger$ \\
\hline
log$_{10}$(L$_{\rm{FIR}}$/L$_\odot$)$^a$ & 13.22$\pm$0.07 & 13.22$\pm$0.12 & 12.91$\pm$0.15 & 13.61$\pm$0.03 & $<$13.36 & 12.47 \\
SFR / M$_\odot$ yr$^{-1}$$^a$ & 2500$\pm$500 & 2500$\pm$900 & 1200$\pm$400 & 6100$\pm$400 & $<$3400 & 270 \\
log$_{10}$(M$_{\rm{dust}}$/M$_\odot$)$^a$ & 8.67$\pm^{0.07}_{0.08}$ & 8.68$\pm^{0.12}_{0.17}$ & 8.40$\pm^{0.11}_{0.15}$ & 9.08$\pm^{0.03}_{0.03}$ & -- & 8.89$\pm^{0.21}_{0.35}$ \\
M$_{\rm{gas}}^{\rm{dust}}$ / $\times$10$^{10}$ M$_\odot$$^b$ & 4.3$\pm$0.9 & 4.4$\pm$1.8 & 2.3$\pm$0.9 & 11$\pm$0.8 & -- & 7.0$\pm^{3.7}_{3.8}$ \\  
L$^{\prime}_{\rm{CO(1-0)}}$ / $\times$10$^{10}$ K km s$^{-1}$ pc$^2$ & 6.5$\pm$0.3 & 4.2$\pm$0.8 & 1.7$\pm$0.1 & 4.1$\pm$0.2 & 6.4$\pm$1.1 & 2.4$\pm$0.5 \\
M$_{\rm{gas}}^{\rm{CO}}$ / $\times$10$^{10}$ M$_\odot$$^c$ & 5.2$\pm$0.3 & 3.4$\pm$0.7 & 1.4$\pm$0.1 & 3.3$\pm$0.2 & 5.1$\pm$0.9 & 1.9$\pm$0.4 \\
Gas-to-dust Ratio$^d$ & 110 & 70 & 55 & 28 & -- & 24 \\
$\alpha_{\rm{CO}}$ / M$_\odot$ (K km s$^{-1}$ pc$^2$)$^{-1}$$^e$ & 0.7 & 1.0 & 1.3 & 2.6 & -- & 2.9 \\
$\Delta$FWHM$_{\rm{CO}}$$^f$ / km s$^{-1}$ & 520$\pm$40 & 700$\pm$120 & 190$\pm$20 & 350$\pm$50 & 840$\pm^{1000}_{350}$ & 550$\pm$200 \\
M$_{\rm{dyn}}$ sin$^2$(i) / $\times$10$^{11}$ M$_\odot$$^g$ & 0.4-3.1 & 0.8-5.7 & 0.07-0.5 & 0.3-1.8 & 0.6 & 4.5 \\
\hline 
\end{tabular}
\end{center}
\begin{small}
\begin{flushleft}
$^\ast$This Work; $^+$\citet{Feruglio:14}; $^\dagger$\citet{Brusa:15} \\ 

(a) Assuming the average of T=47K, $\beta$=1.6 and T=41K, $\beta$=1.95 for the four quasars in this work as well as ULASJ1539+0557; For ULASJ1002+0137 the FIR luminosity and SFR are taken directly from \citet{Brusa:15}. \\  
(b) Assuming gas-to-dust ratio of 91 \citep{Sandstrom:13}; \\
(c) Assuming $\alpha_{\rm{CO}}=0.8$ M$_\odot$ (K km s$^{-1}$ pc$^2$)$^{-1}$; \\
(d) Assuming M$_{\rm{gas}}$=M$_{\rm{gas}}^{\rm{CO}}$; \\ 
(e) Assuming M$_{\rm{gas}}$=M$_{\rm{gas}}^{\rm{dust}}$; \\
(f) Corrected for instrumental resolution due to Hanning smoothing of spectra; \\
(g) Full range in values quoted using the different estimators detailed in Section \ref{sec:dynamics} for the four quasars in this work. For ULASJ1539+0557 and ULASJ1002+0137, the dynamical masses are taken directly from \citet{Feruglio:14} and \citet{Brusa:15}. \\
\end{flushleft}
\end{small}
\end{table*}
\end{landscape}

\subsection{$^{12}$CO(3-2) Line Luminosities and Molecular Gas Masses}

\label{sec:CO}

All four quasars are also detected in the $^{12}$CO(3-2) line. Most of these quasars were not previously known to be millimetre bright and were identified for follow-up observations based purely on their near infrared colours and implied dust extinctions. This strongly suggests that the near-infrared selection is successfully identifying broad line quasars in gas rich, highly star forming host galaxies at $z\sim2-3$.

\begin{figure*}
\begin{center}
\vspace{-2cm}
\begin{tabular}{cc}
\vspace{-5cm}
 \includegraphics[scale=0.35]{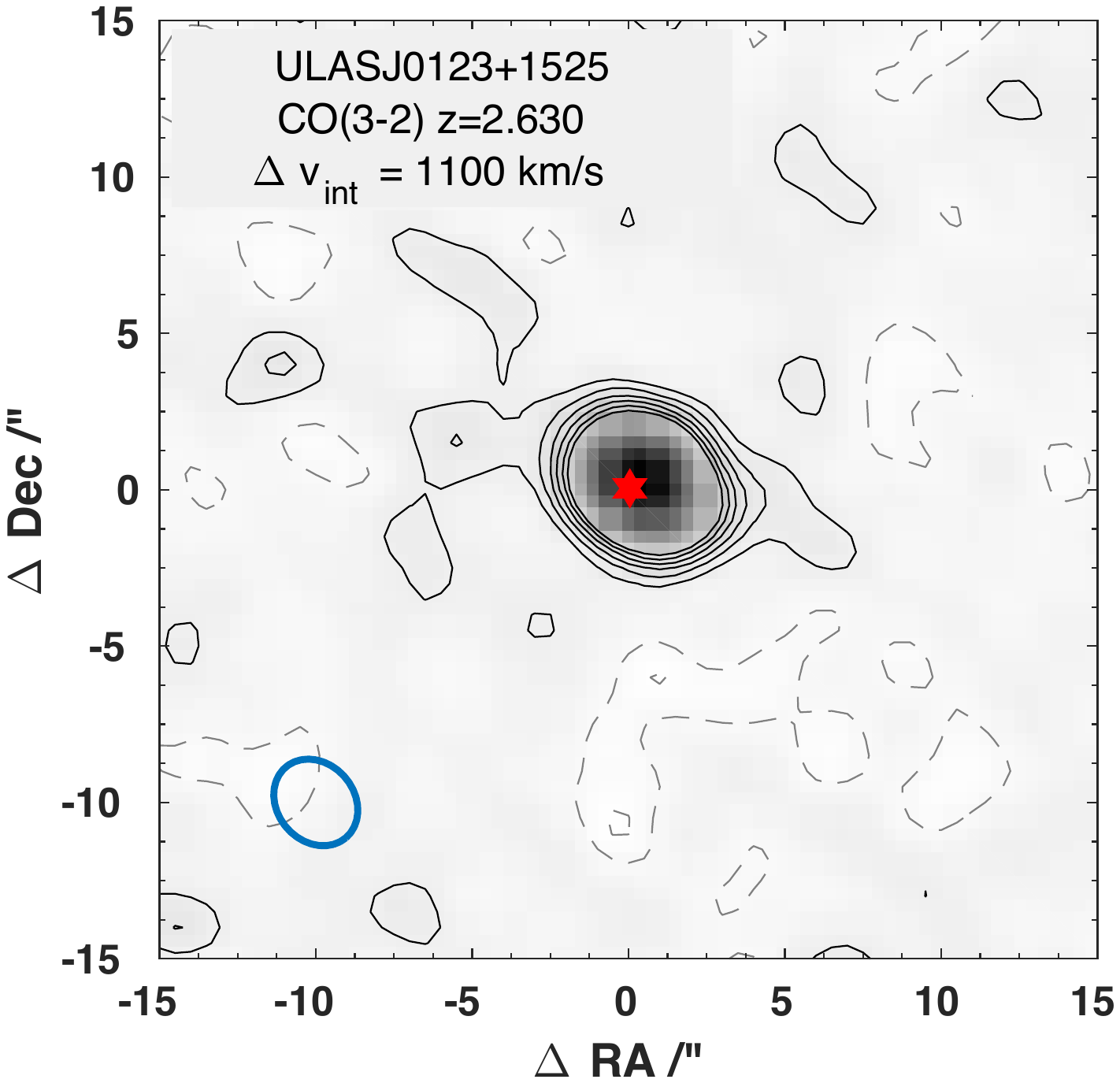} & \includegraphics[scale=0.35]{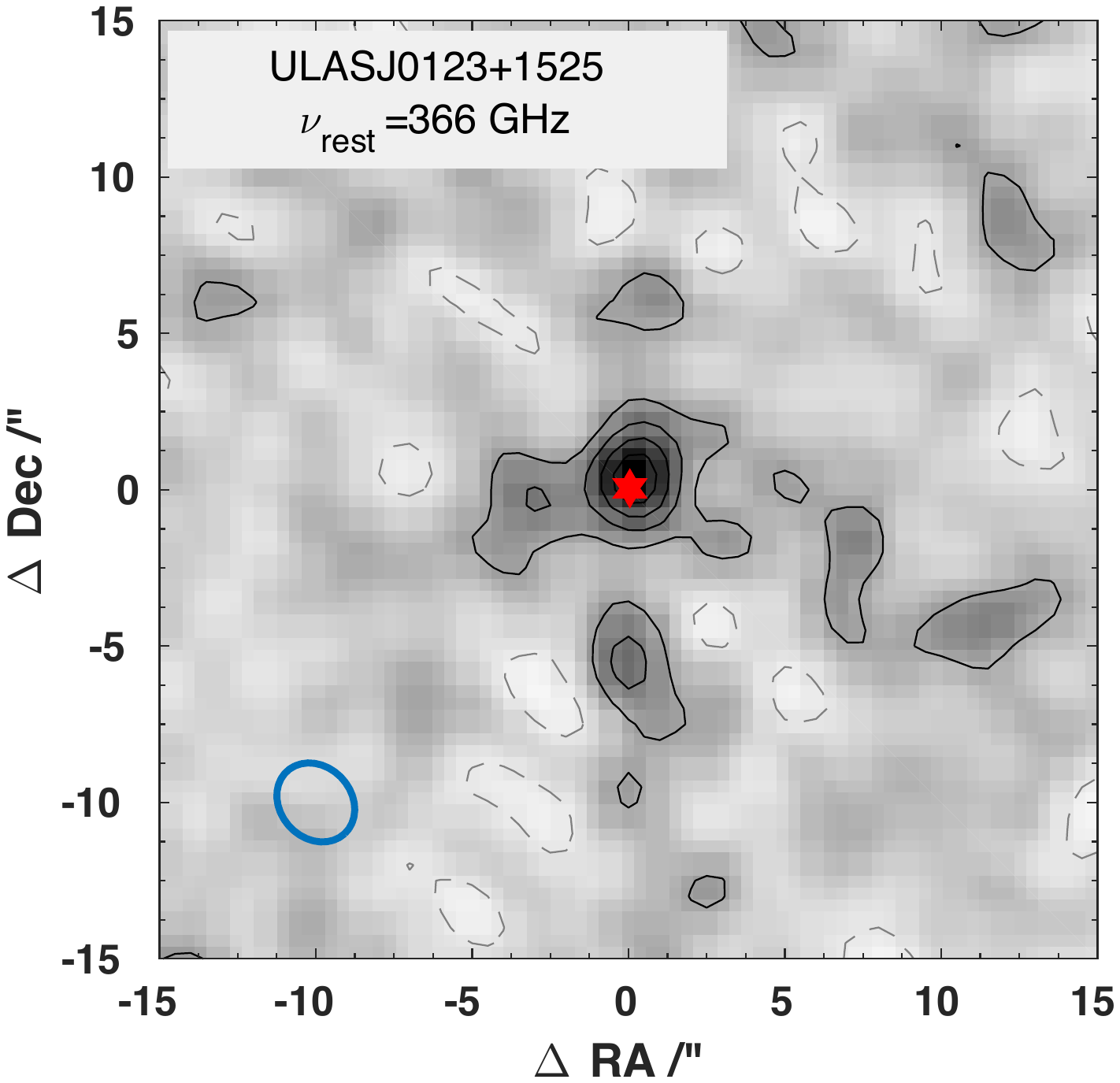} \\
 \vspace{-5cm}
  \includegraphics[scale=0.35]{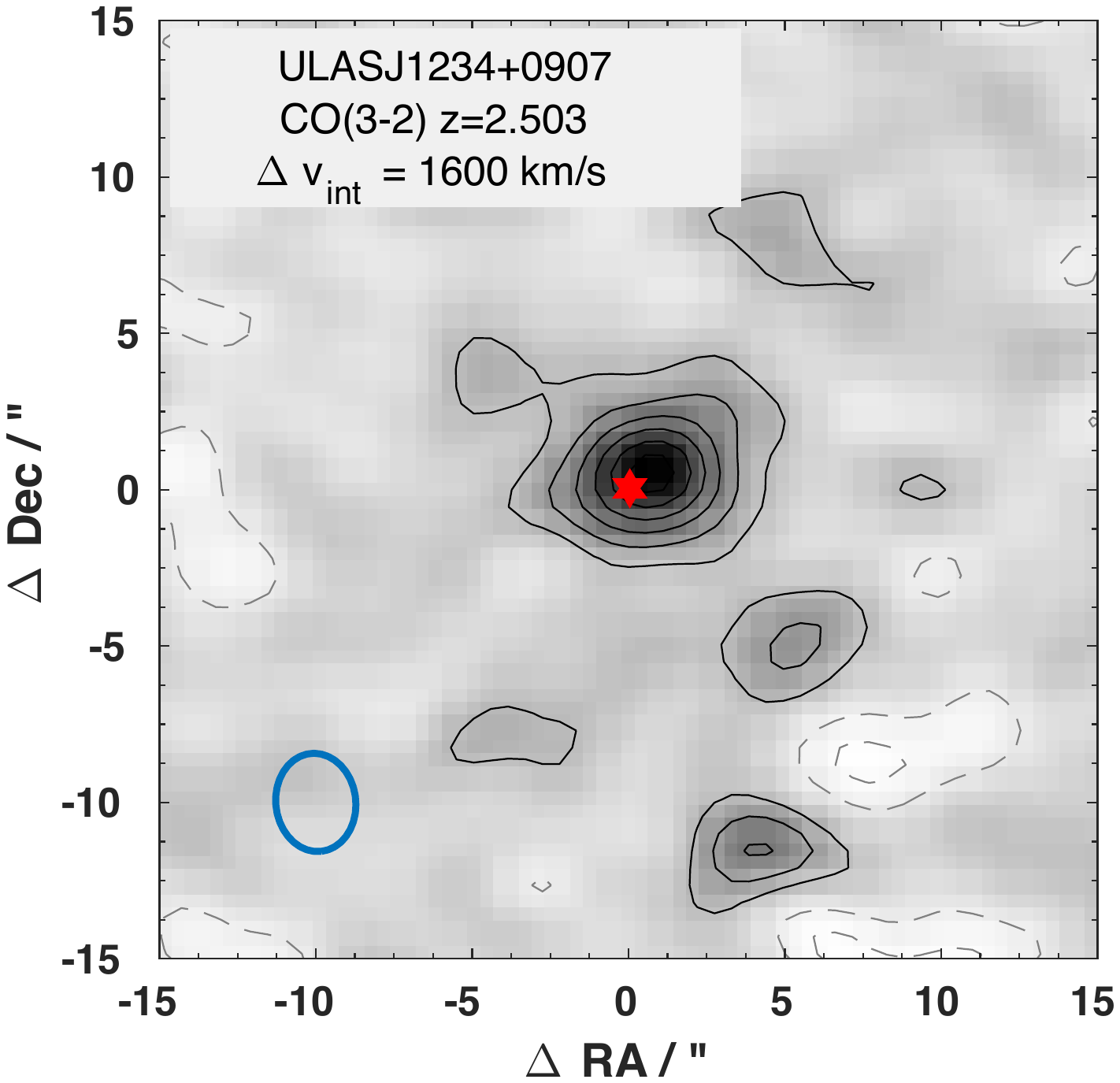} & \includegraphics[scale=0.35]{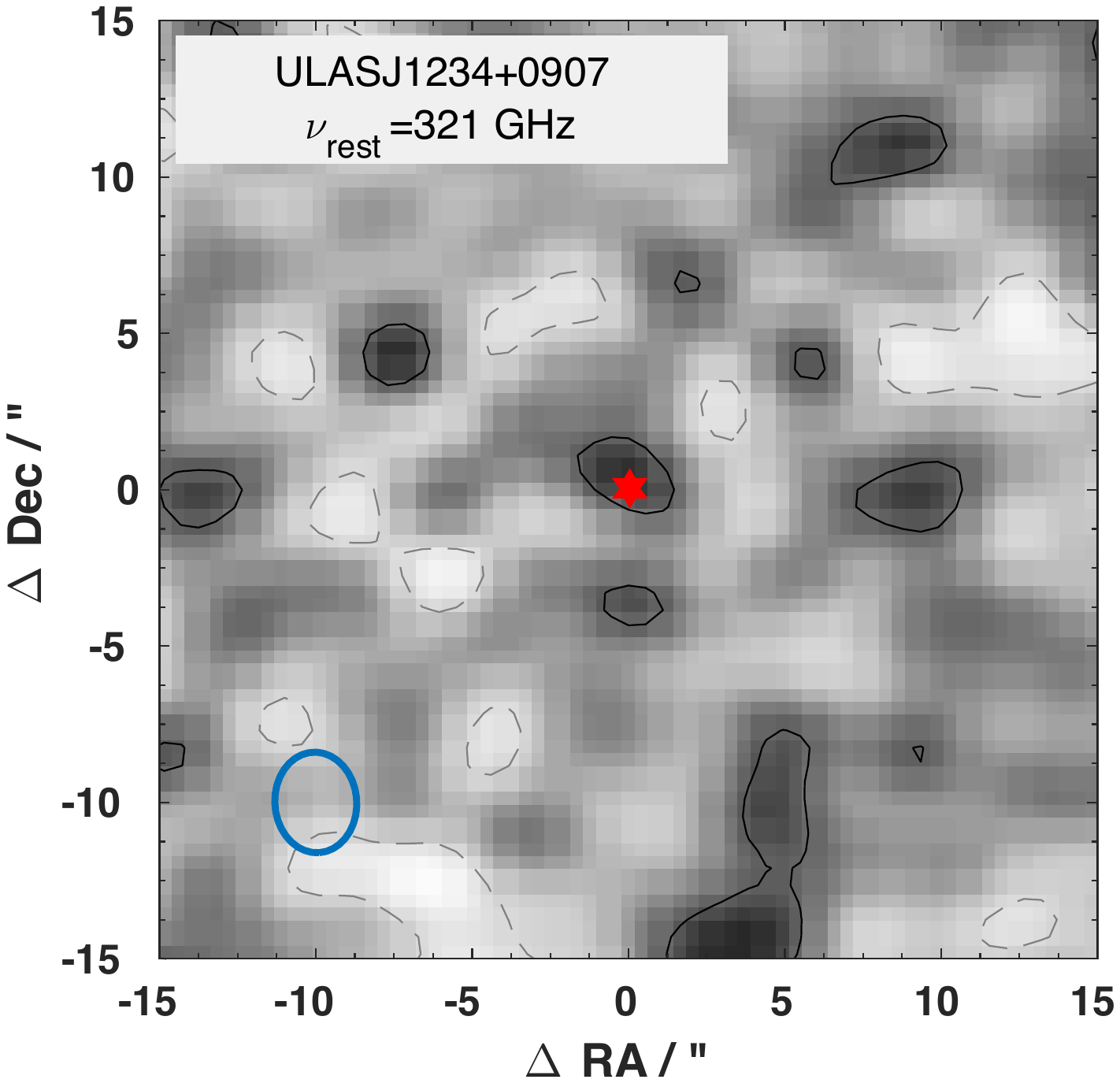} \\
  \vspace{-5cm}
\includegraphics[scale=0.35]{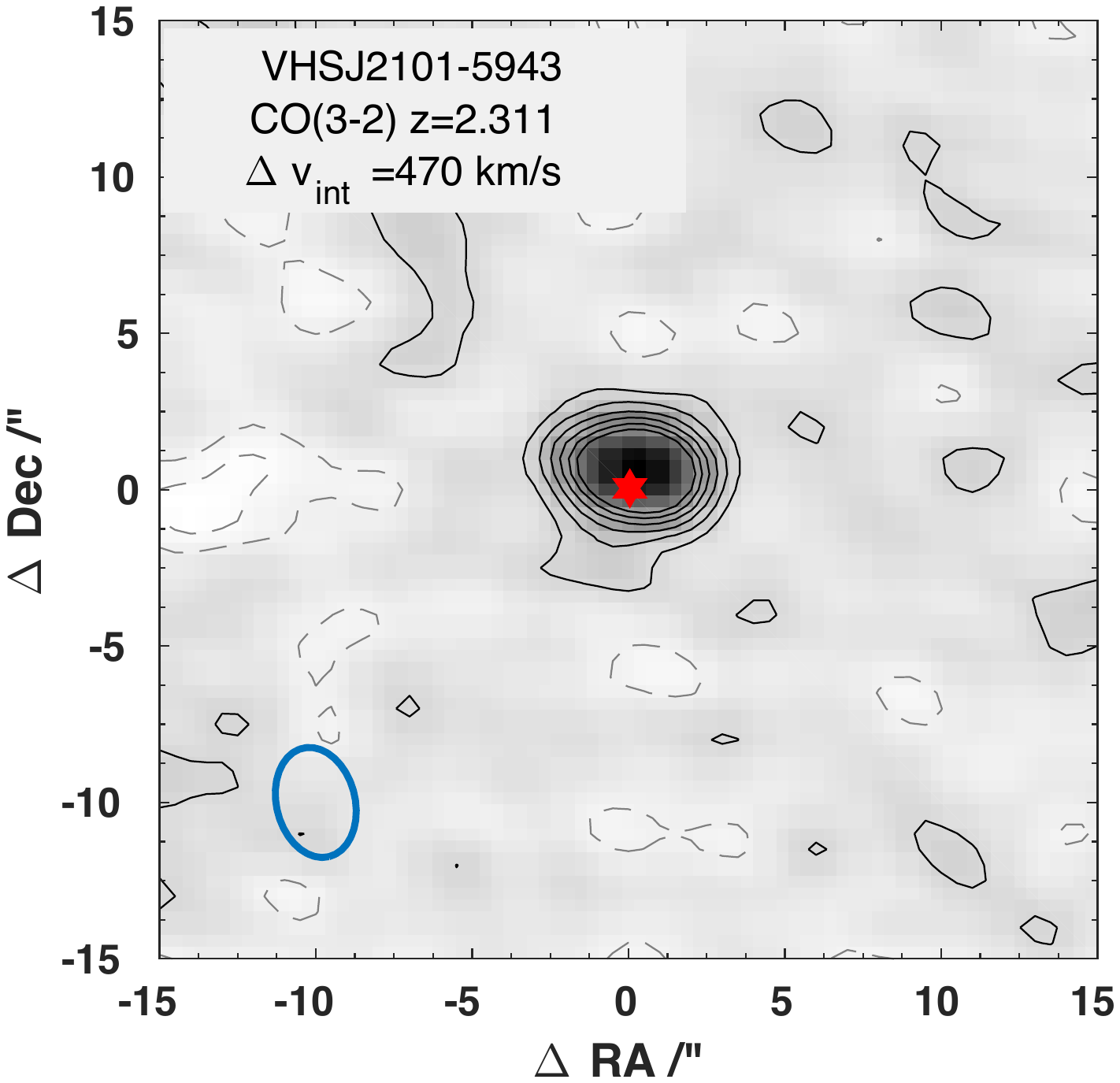} & \includegraphics[scale=0.35]{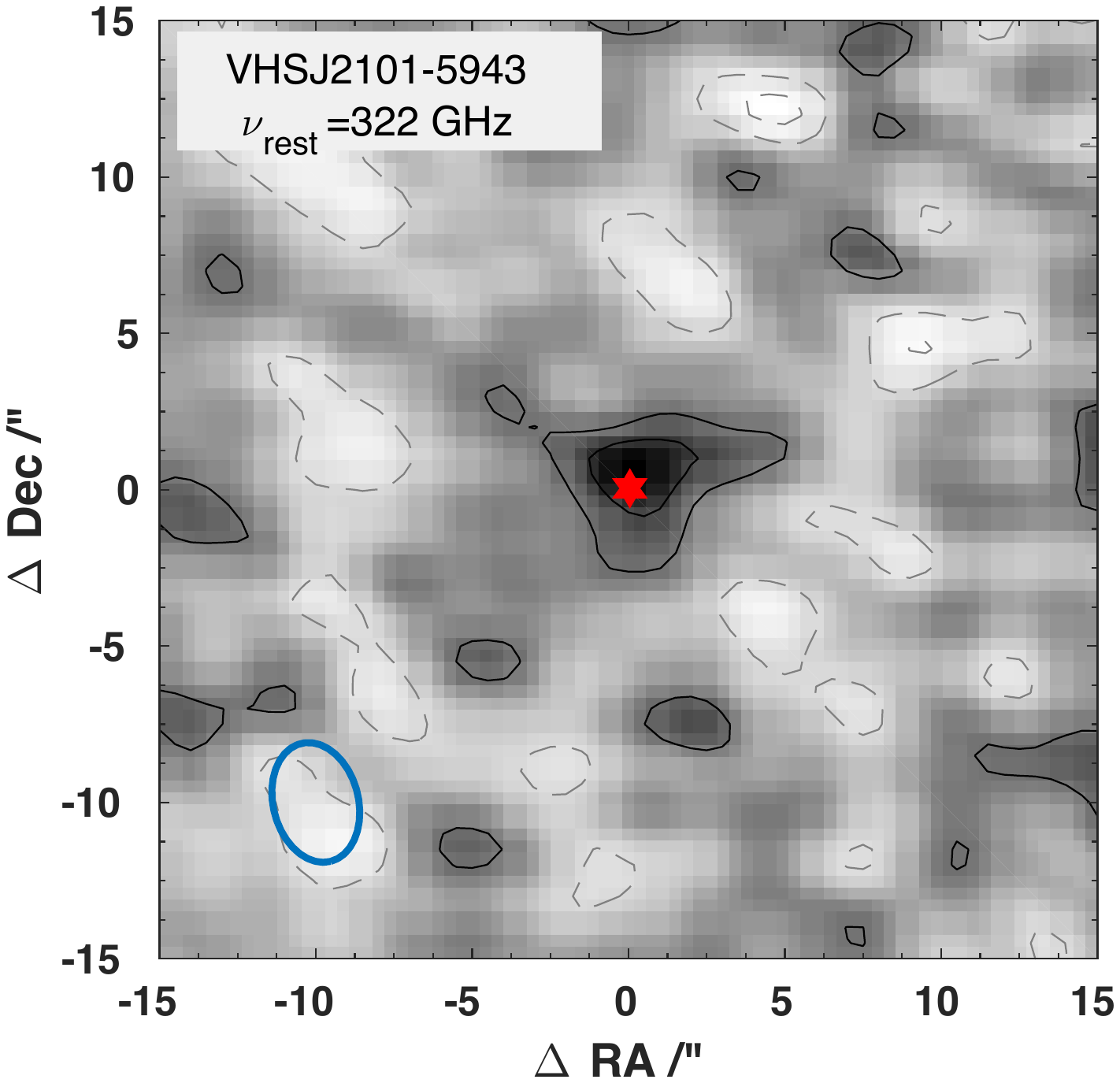} \\
\vspace{-2cm}
 \includegraphics[scale=0.35]{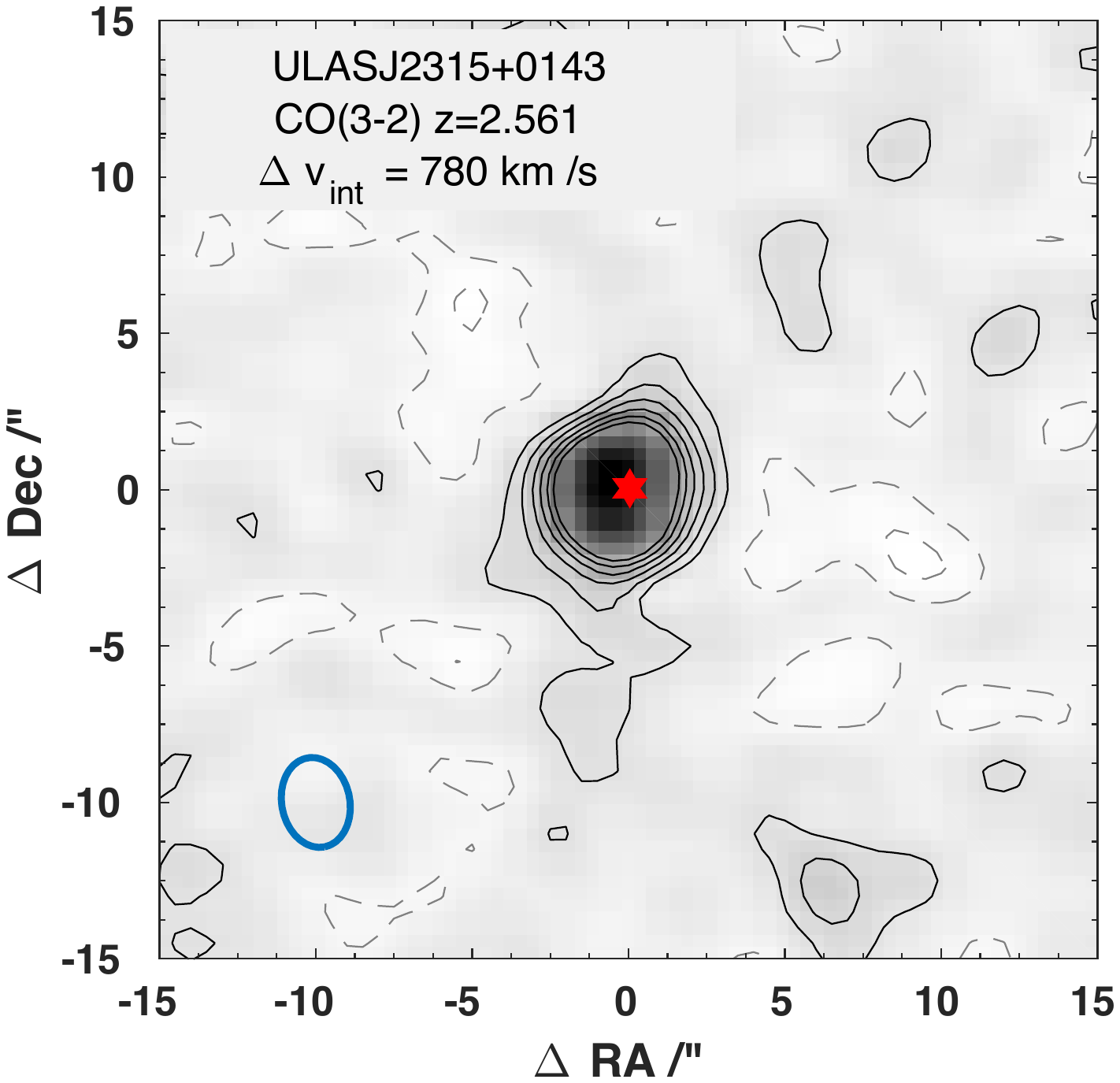} & \includegraphics[scale=0.35]{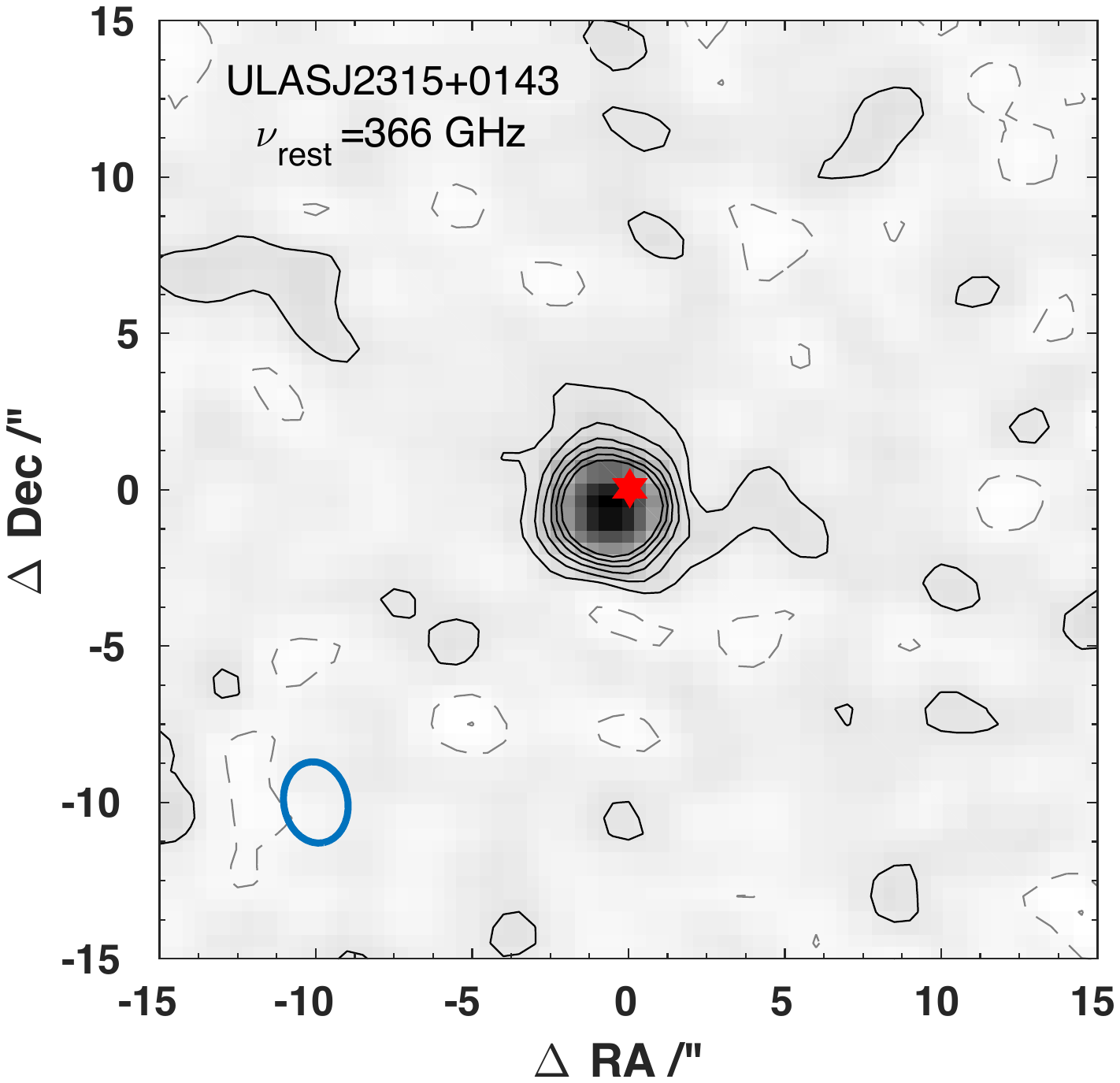} \\
\end{tabular}
\caption{$^{12}$CO(3-2) (left) and dust continuum (right) maps for all four quasars in our sample. Contours start at $\pm$1.5$\sigma$ and are shown in intervals of 1.5$\sigma$. Solid contours denote regions of positive flux while dashed contours denote regions of negative flux. The star marks the position of the quasar from the near infrared observations. The beam size is indicated in the bottom left corner of the map.}
\label{fig:maps}
\end{center}
\end{figure*}

\begin{figure*}
\begin{center}
\begin{tabular}{cc}
\vspace{-3cm}
\large{ULASJ0123$+$1525 (z=2.630)} & \large{ULASJ1234$+$0907 (z=2.503)} \\
\vspace{-3cm}
\includegraphics[scale=0.4]{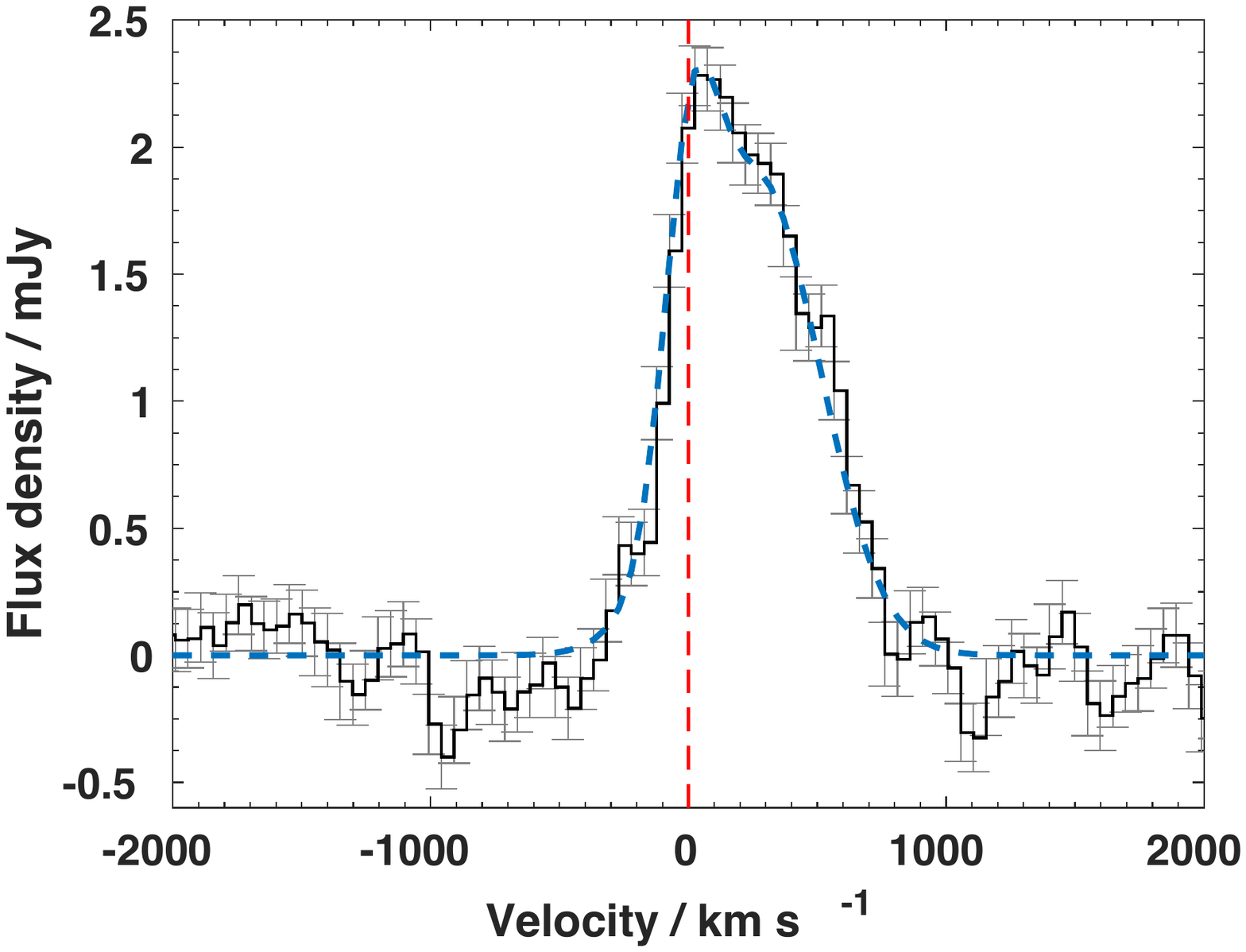} & \includegraphics[scale=0.4]{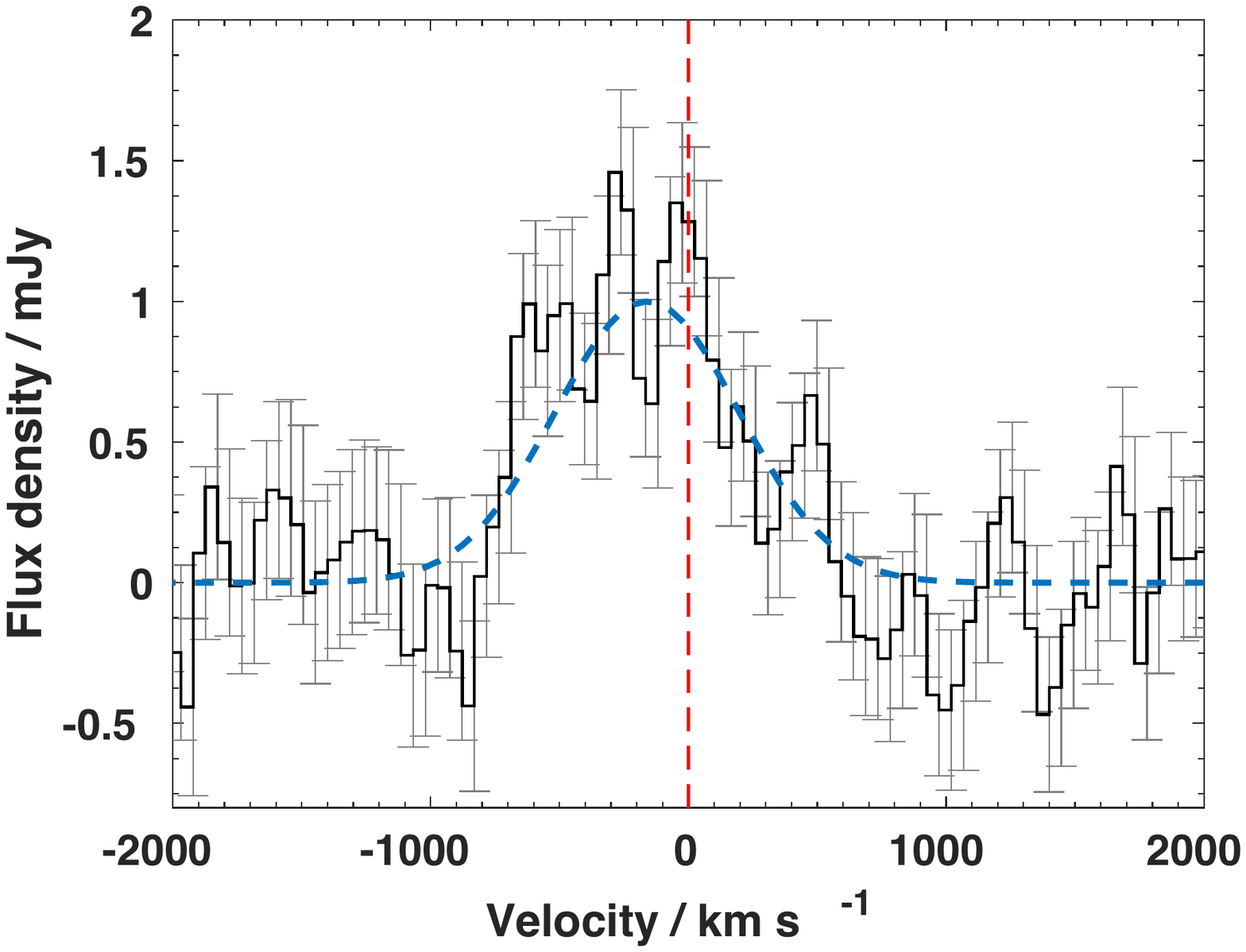} \\
\vspace{-3cm}
\large{VHSJ2101$-$5943 (z=2.311)} & \large{ULASJ2315$+$0143 (z=2.561)} \\
 \includegraphics[scale=0.4]{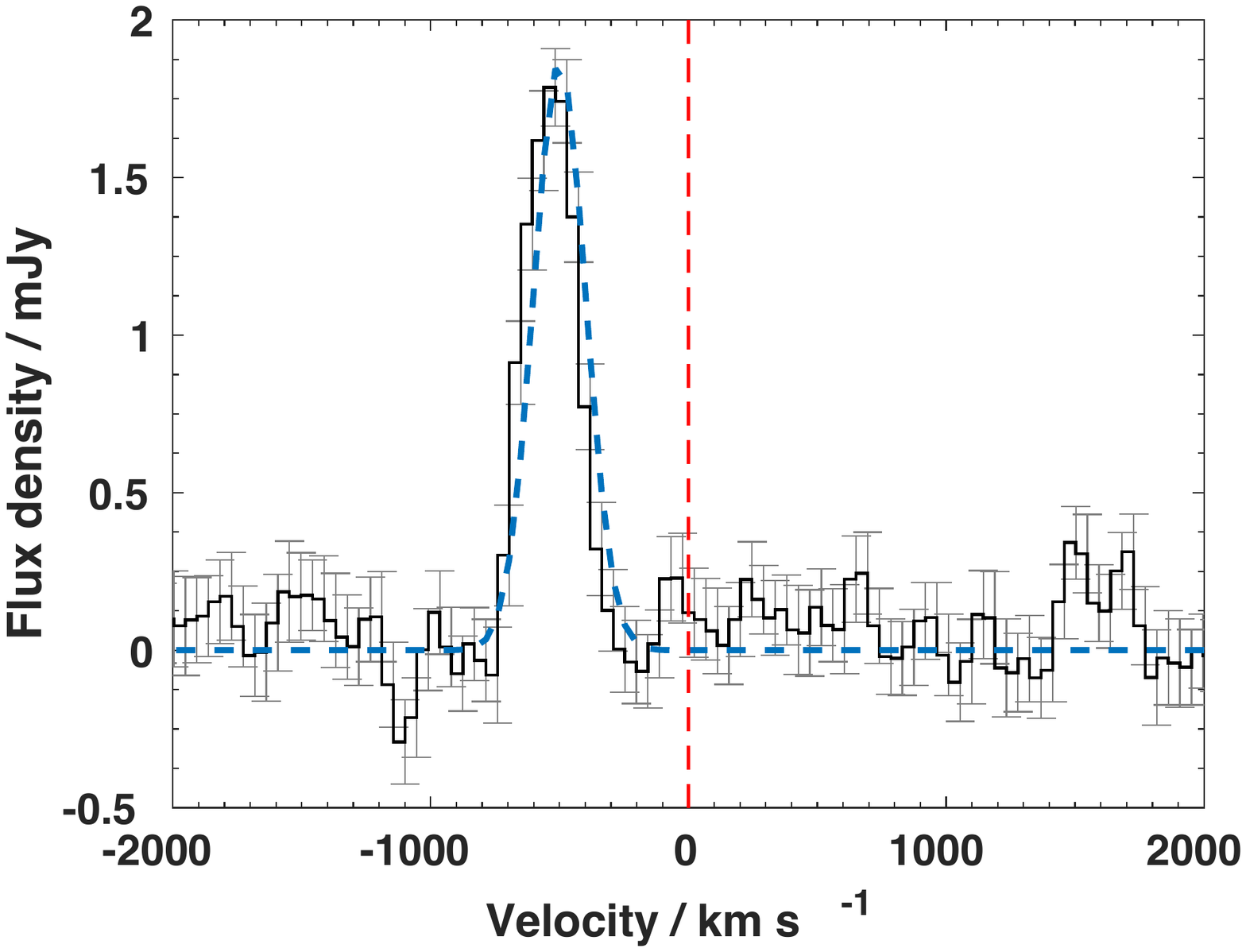} & \includegraphics[scale=0.4]{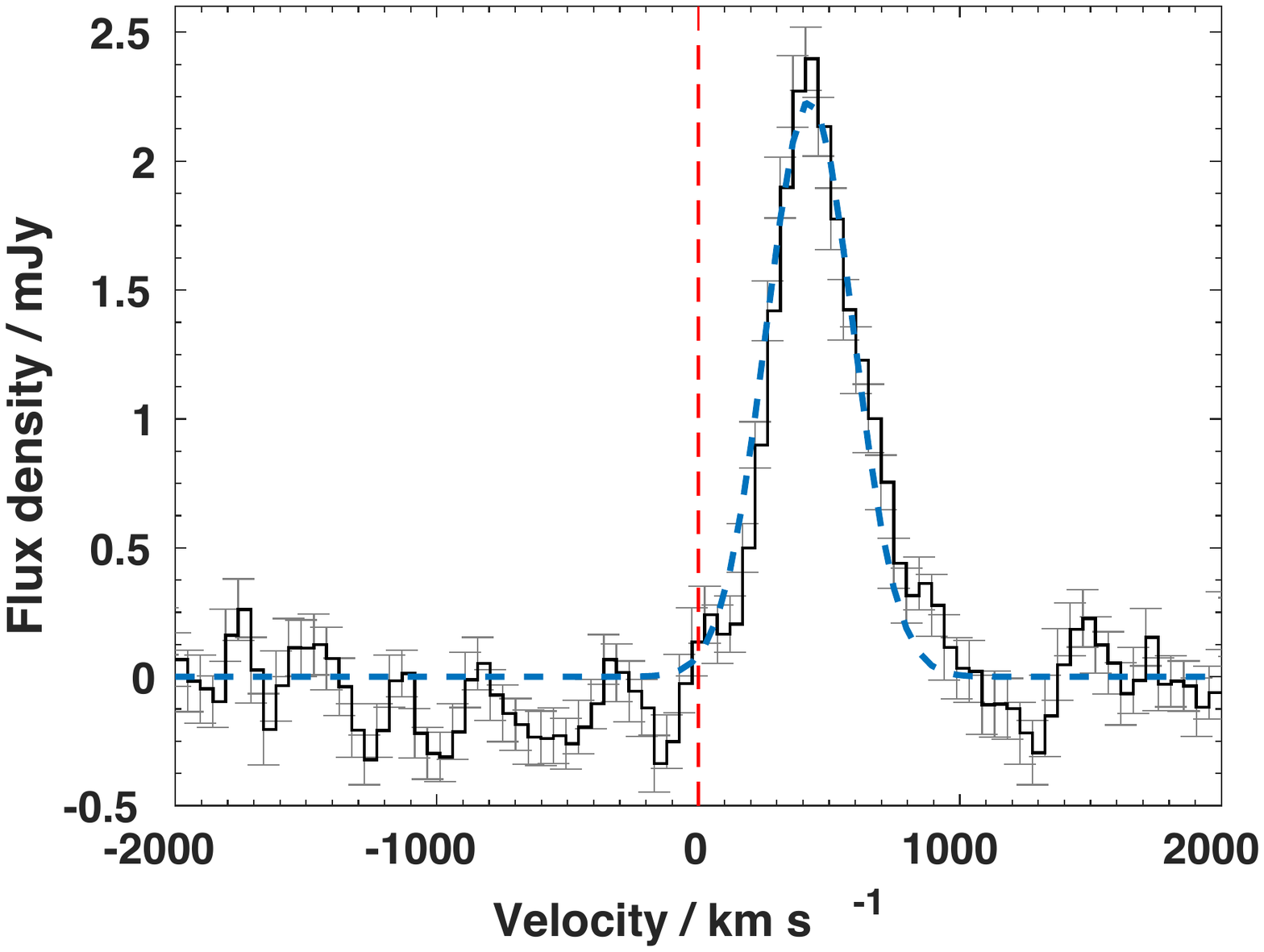} \\
\end{tabular}
\vspace{-3cm}
\caption{$^{12}$CO(3-2) spectra for all four quasars in our sample. The error bars denote the 1$\sigma$ uncertainties on the measured flux. The spectra are Hanning smoothed straight out of the ALMA pipeline. Dashed lines show the best fit Gaussian profile fit to these data. In the case of ULASJ0123$+$1525 two Gaussians have been used to fit the data. The velocities are relative to the H$\alpha$ redshifts from B12 and B15 and the vertical dashed line marks zero velocity.}
\label{fig:spec}
\end{center}
\end{figure*}

Line maps extracted over the full-width-half-maximum (FWHM) velocity interval centred on the line can be seen in Fig. \ref{fig:maps} together with the corresponding CO spectra in Fig. \ref{fig:spec}. We checked by fitting a 2-dimensional (2-D) Gaussian to the line maps that the centroid of the CO line emission was always within 0.5-1 arcsec of the centroid of the quasar emission as measured from the near infrared images in B12 and B15. The largest offset is seen in ULASJ2315+0143 where the peak CO emission lies 1 arcsec from the quasar emission. ULASJ2315$+$0143 is spatially resolved in our data and is discussed further below in Section \ref{sec:dynamics} and \ref{sec:2315}. All CO line properties quoted for ULASJ2315+0143 correspond to those derived from the spatially integrated spectrum of this source. 

Our observations of the dust continuum as well as the CO(3-2) line reveal a range in the CO(3-2)-to-FIR luminosities in the sample. While ULASJ0123+1525 has L$_{\rm{FIR}}$/L$_{\rm{CO(3-2)}}$=2.5$\times$10$^5$, in ULASJ2315+0143 this ratio is 2.2$\times$10$^6$ - i.e. almost an order of magnitude larger. Seen another way, the CO equivalent widths in the sample also cover a large range from $\sim$3.5$\times$10$^{3}$ km s$^{-1}$ in ULASJ2315+0143 to $\sim$1.4$\times$10$^{4}$ km s$^{-1}$ in ULASJ0123+1525. For ULIRG type systems, the CO(3-2) line typically comprises $\sim$25 per cent of the $\sim$850$\mu$m flux within a $\sim$3$\times$10$^{4}$ km s$^{-1}$ bandwidth, which corresponds to an equivalent width of the CO(3-2) line of $\sim$1$\times$10$^{4}$ km s$^{-1}$ \citep{Seaquist:04}. The range in CO-to-FIR luminosities and equivalent widths seen in our reddened quasars may indicate a range in ISM properties in the sample and further observations of other molecular lines and CO transitions would help constrain the excitation conditions of the gas in these systems. 

The dust and gas detections also allow us to calculate molecular gas masses in two different ways. Recently there have been several studies in the literature advocating the use of dust masses as a proxy for the gas mass assuming a metallicity dependent gas-to-dust ratio (e.g. \citealt{Scoville:14}). By assuming a gas-to-dust ratio of 91$\pm^{60}_{36}$ from observations of nearby galaxies \citep{Sandstrom:13}, we can convert the dust masses in Table \ref{tab:derived} to gas masses and, using the CO line luminosity, we can therefore estimate the CO-to-H$_2$ conversion factor, $\alpha_{\rm{CO}}$. These values can be seen in Table \ref{tab:derived} and we find that a range of $\alpha_{\rm{CO}}$ values are derived for the sample all the way from $\sim$0.7 M$_\odot$ (K km s$^{-1}$ pc$^2$)$^{-1}$ appropriate for nuclear starbursts and quasars to $\sim$3 (K km s$^{-1}$ pc$^2$)$^{-1}$ in ULASJ2315+0143, appropriate for disk like galaxies like our own Milky Way as well as high redshift star forming disks. It is interesting to note that ULASJ2315+0143 is also the source with the most extended gas emission. However, we caution that these values are highly dependent on the assumptions made regarding the dust SED of the quasars in Section \ref{sec:dust}.

Gas masses can also be estimated directly from the CO line luminosities, which are also given in Table \ref{tab:derived}. We assume an excitation ratio between the $^{12}$CO(3-2) and $^{12}$CO(1-0) line of $r_{32/10}$=0.8, which is intermediate between the typical values for SMGs and optical quasars from \citet{Carilli:13} and consistent with our interpretation that reddened quasars are transitioning from an SMG-like to an optical quasar-like phase. The corresponding CO(1-0) line luminosities can be seen in Table \ref{tab:derived} and these are converted to gas masses assuming $\alpha_{\rm{CO}}=0.8$ (K km s$^{-1}$ pc$^2$)$^{-1}$ appropriate for nuclear starbursts and quasars. The resultant gas-to-dust ratios presented in Table \ref{tab:derived} range from values of $\sim$100, similar to the values in nearby galaxies, to $\sim$30-50, consistent with other high redshift star forming galaxies \citep{Aravena:16}, but are once again dependant on the exact choice of dust SED parameters.

\subsection{$^{12}$CO(3-2) Kinematics and Dynamical Masses}

\label{sec:dynamics}

\begin{figure}
\begin{center}
\vspace{-3cm}
\hspace{-1cm}
\includegraphics[scale=0.45]{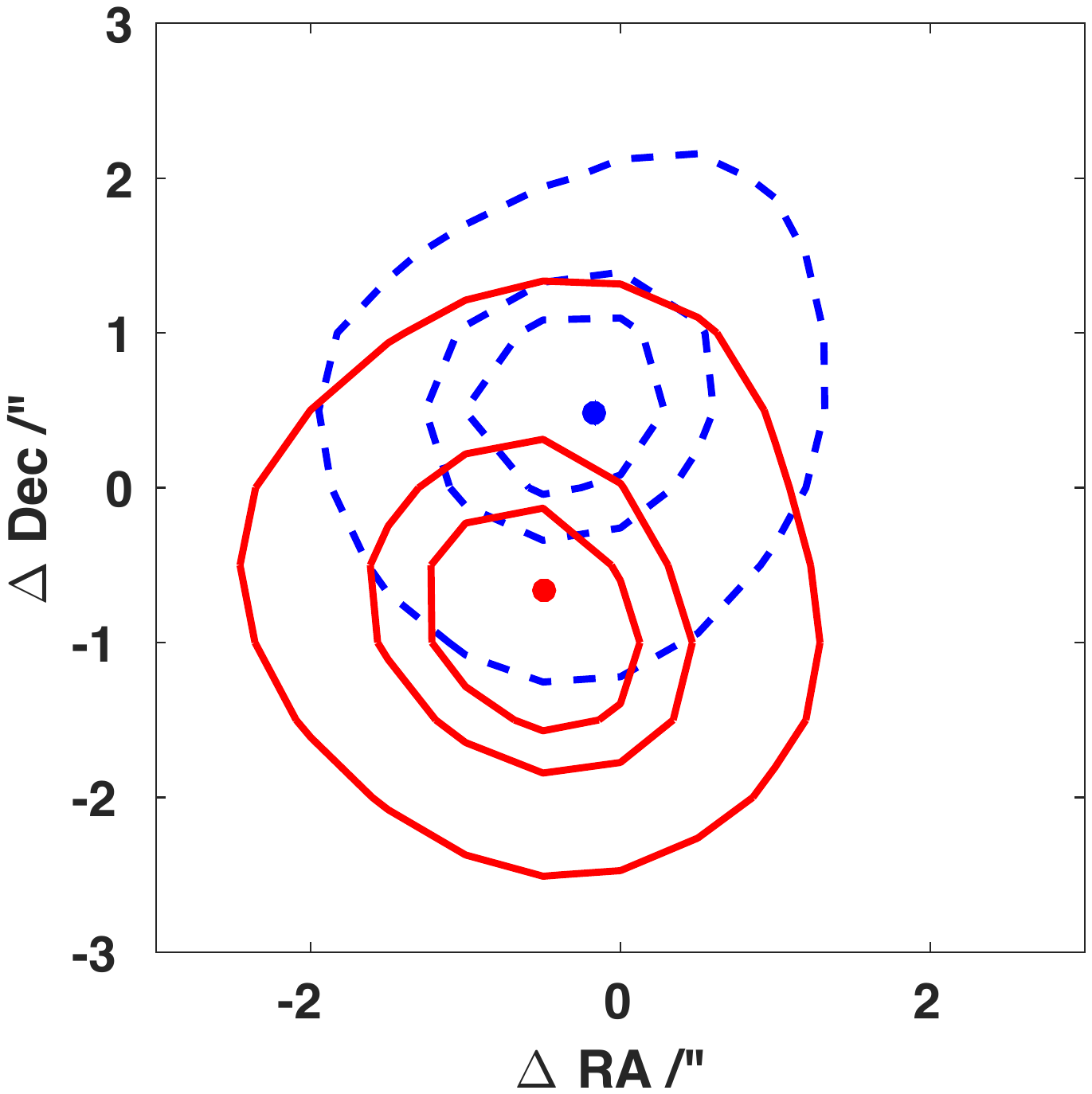} \\
\vspace{-3cm}
\caption{CO line maps for ULASJ2315+0143 extracted using the blue and red sides of the line separately. Given the high signal-to-noise (S/N) of these data, the typical centroiding errors for the red and blue wings are $\sim$0.07$\arcsec$ and are therefore too small to be visible on this plot. A clear spatial offset is seen between the two maps with the dots marking the centres of the two maps as calculated using a 2-D (in RA, Dec) Gaussian fitting. This suggests that the source is spatially resolved.}
\label{fig:specast}
\end{center}
\end{figure}

Our observations were carried out in low resolution mode corresponding to the most compact ALMA Cycle 3 configuration. However, the gas emission nevertheless appears to be extended in some of our quasars. We tested for evidence for spatially resolved emission using the spectroastrometry method detailed in e.g. \citet{Carniani:13}. Specifically, we constructed separate line maps from the red and blue side of the line and looked for evidence for a spatial offset between the two maps. These line maps can be seen in Fig. \ref{fig:specast} for our most extended source, ULASJ2315+0143. We measure a (1.20$\pm$0.13) arcsec offset between the two maps in Fig. \ref{fig:specast} corresponding to physical scales of (9.6$\pm$1.0) kpc. The dust continuum emission from ULASJ2315+0143 is more compact and spatially unresolved in our data, suggesting that the gas reservoirs extend to larger scales than the cold dust in this galaxy. Using the same method, we find a (0.78$\pm$0.16) arcsec $\equiv$ (6.4$\pm$1.3 kpc) offset between the red and blue emission in VHSJ2101$-$5943 and a (0.58$\pm$0.2) arcsec $\equiv$ (4.7$\pm$1.6 kpc) offset in ULASJ1234+0907. In ULASJ0123+1525 the red and blue wing emission map centroids are separated by (0.04$\pm$0.09) arcsec $\equiv$ (0.3$\pm$0.7 kpc). In all cases (apart from ULASJ2315+0143 which is discussed in Section \ref{sec:2315}), the line emission is therefore spatially unresolved or only marginally resolved. 

We use either a single or double Gaussian to fit the line profiles in Fig. \ref{fig:spec} to characterise the line shape and also compute non-parametric first and second intensity weighted moments to derive line centroids and line widths. The CO redshifts quoted in Table \ref{tab:obs} are derived from these line centroids. The CO-derived redshifts are all within $\pm$250 km s$^{-1}$ of the H$\alpha$ derived redshifts in B12 and B15. Given the broad H$\alpha$ line profiles in these quasars, the typical uncertainties in the H$\alpha$ redshift estimates can be several hundred km s$^{-1}$.

In all cases, apart from ULASJ0123+1525, a single Gaussian gives an acceptable fit to the data with a reduced $\chi^2$ value of $\sim$1. In the case of ULASJ0123+1525, a double Gaussian leads to a considerably better fit to the data. The parametric and non-parametric values for the line centroid and line widths are consistent within the error bars and in Table \ref{tab:derived} we quote the non-parametric widths of the CO emission line. 

Dynamical masses can be estimated from the widths of the CO lines in Table \ref{tab:derived} assuming a source size. We have already seen that in ULASJ2315+0143 the gas emission is spatially resolved with an implied source radius of $\sim$4.8 kpc from the spectroastrometry method. We can use the relation in \citet{Gnerucci:11} to derive the corresponding dynamical mass:

\begin{equation}
M_{\rm{dyn}} sin^2(i) = (1.0\pm0.1) \times 2.3 \times 10^9 M_\odot \left(\frac{FWHM}{100 \rm{km s}^{-1}}\right)^2 \left(\frac{r_{\rm{spec}}}{1 \rm{kpc}}\right)
\label{eq:mdyn}
\end{equation}

\noindent This assumes that the gas emission comes from a rotating disk with the red and blue images shown in Fig. \ref{fig:specast} tracing different sides of the disk due to rotation. In such an estimator, $r_{\rm{spec}}$ does not correspond to a conventional size but is rather the average distance between the redshifted and blueshifted gas weighted by the surface brightness of the CO line. We leave the dependence on inclination explicit in the derivations of dynamical masses. This gives M$_{\rm{dyn}}sin^2(i)$=1.5$\times10^{11}$M$_\odot$ for ULASJ2315+0143. Assuming a typical R=4 kpc for the other three quasars, the dynamical masses are (1.3, 2.3 and 0.2)$\times10^{11}$M$_\odot$ for ULASJ0123+1525, ULASJ1234+0907 and VHSJ2101-5943 respectively. For completeness, we also estimate the dynamical masses using two other estimators: 1) the virial estimator: M$_{\rm{dyn}}$=2.82$\times$10$^5$ $\times {\rm{FWHM}}^2 \times R$ and the rotating disk estimator \citep{Neri:03}: M$_{\rm{dyn}}$sin$^2$(i)=4$\times$10$^4$ $\times {\rm{FWHM}}^2 \times R$. The values quoted in Table \ref{tab:derived} encompass the full range in dynamical masses from these three estimators. 

\subsection{Notes on Individual Objects}

\subsubsection{ULASJ1234+0907: A Significant Overdensity of Millimetre Bright Galaxies}

\label{sec:1234}

\begin{figure*}
\begin{center}
\vspace{-3cm}
\begin{tabular}{cc}
\vspace{-5.5cm}
\includegraphics[scale=0.4]{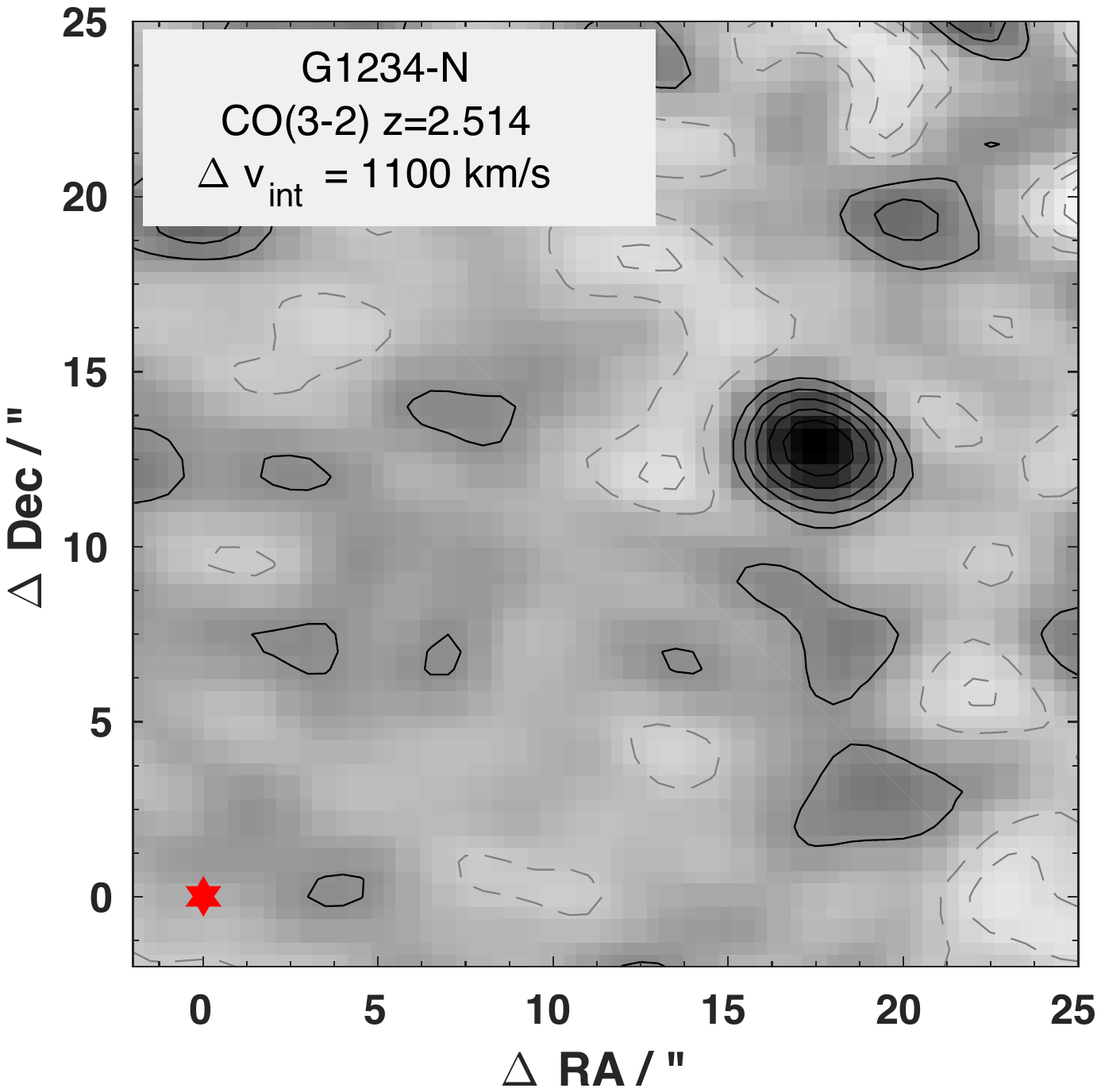} & \includegraphics[scale=0.4]{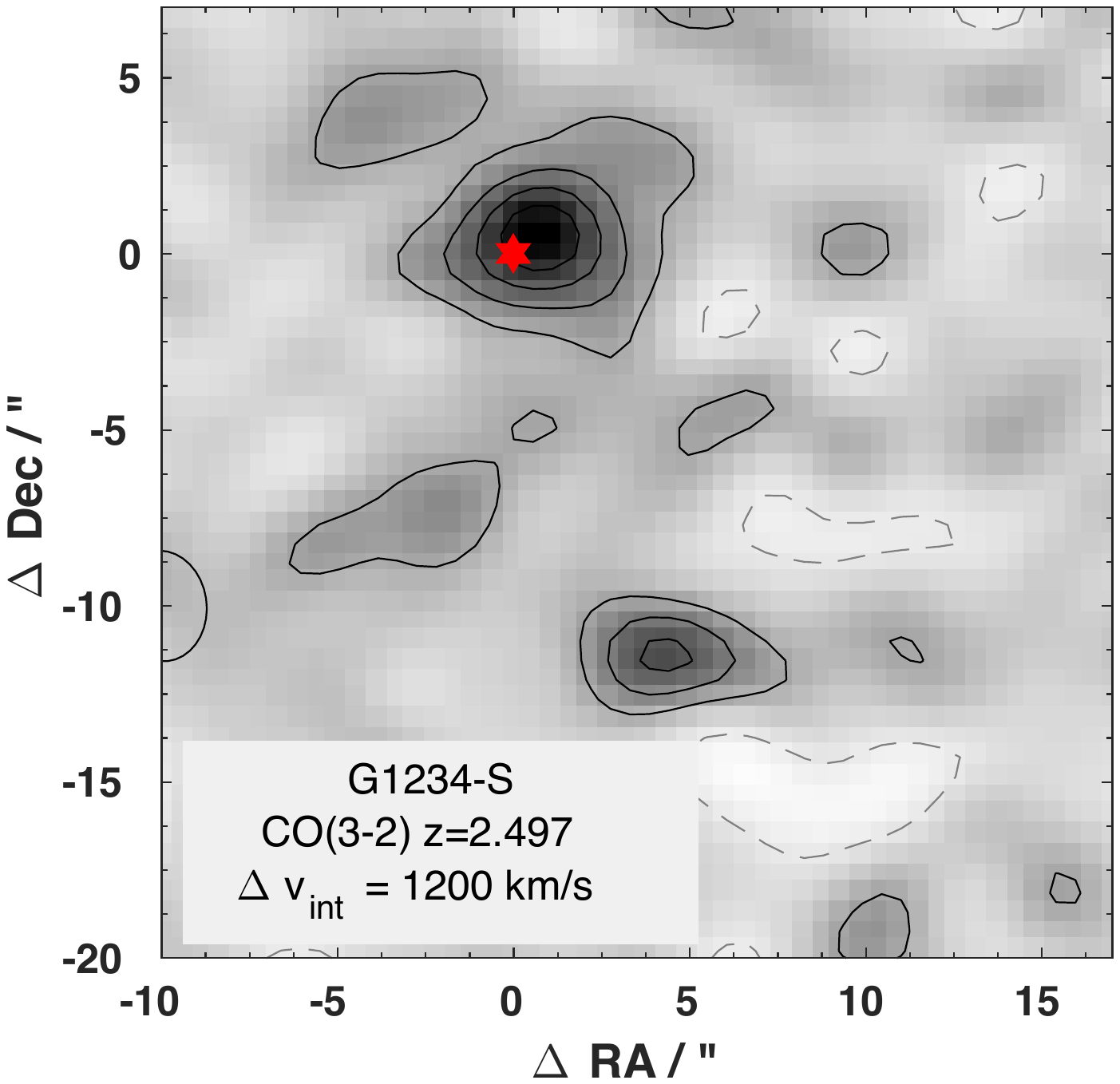} \\
\vspace{-0.cm}
\includegraphics[scale=0.4]{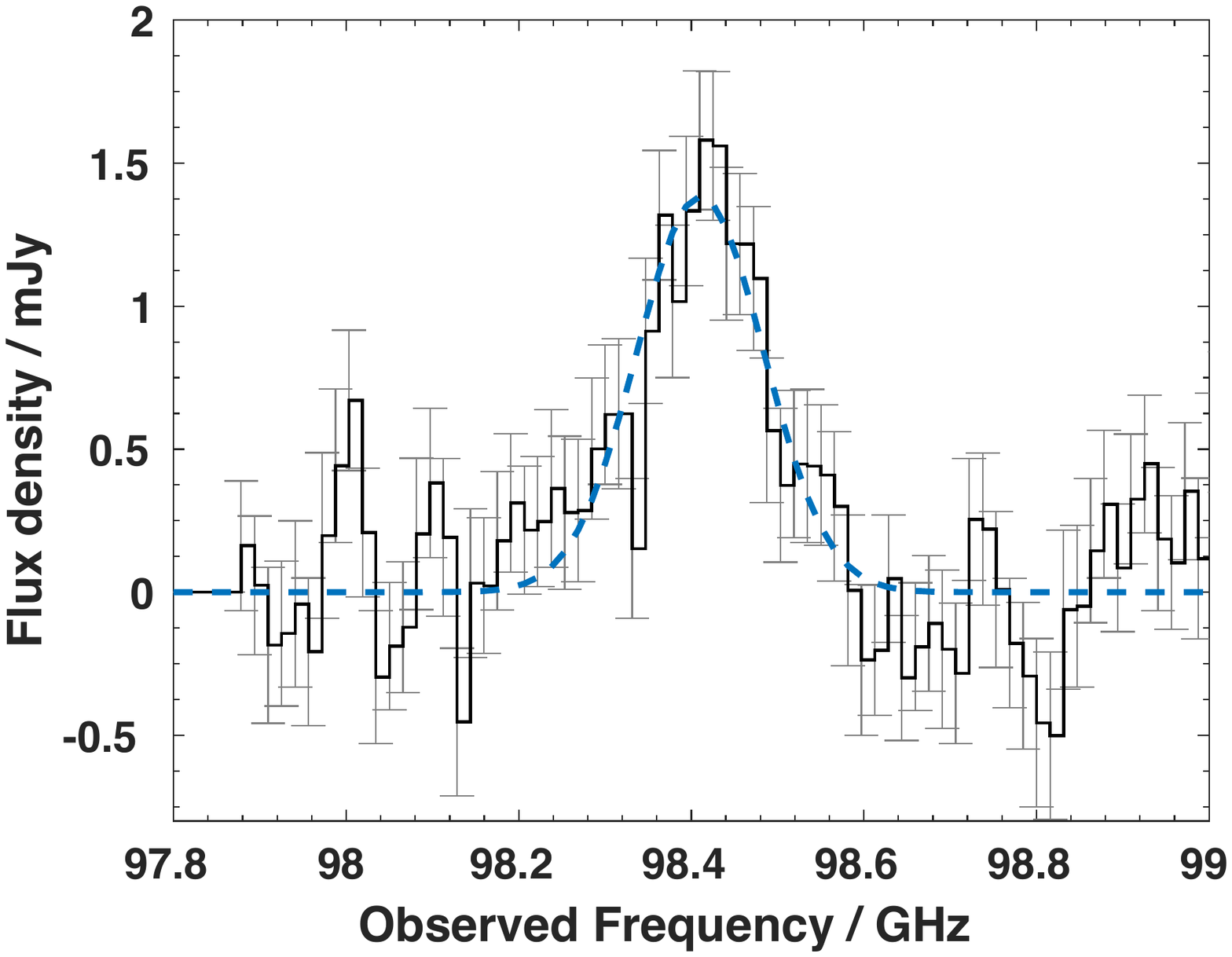} & \includegraphics[scale=0.4]{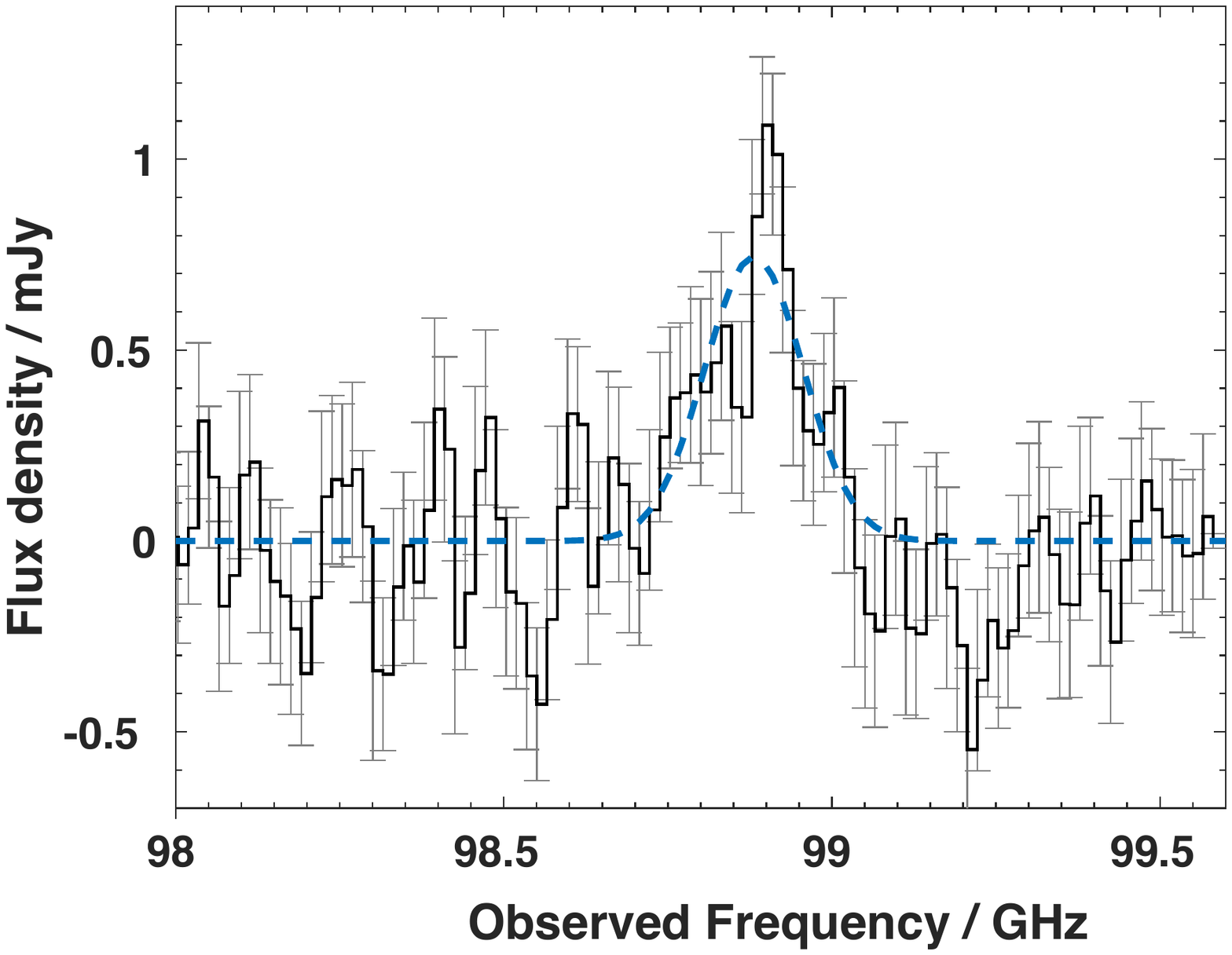} \\
\end{tabular}
\vspace{-3cm}
\caption{\textit{\underline{Top:}} CO line maps constructed by summing the channels containing the CO line in G1234-N (left) and G1234-S (right). The offsets in RA and Dec are relative to the position of the quasar (the red star), ULASJ1234$+$0907 which marks the phase center. G1234-N lies north-east of the quasar while G1234-S lies south of the quasar. The G1234-S channel map overlaps in frequency with the CO line in ULASJ1234+0907 and both sources are therefore visible in the map. \textit{\underline{Bottom:}} $^{12}$CO(3-2) spectrum for G1234-N (left) and G1234-S (right) with the dashed line showing the best fit Gaussian fit to the line profile. The error bars denote the 1$\sigma$ uncertainties on the flux measurements.}
\label{fig:1234NS}
\end{center}
\end{figure*}

Given that our high luminosity, massive quasars are likely to be the progenitors of the most massive galaxies seen in the Universe today, they may be expected to reside in massive halos. Significant overdensities of galaxies around the quasar may therefore be expected even at $z\sim2-3$. The ALMA field-of-view at these frequencies (FWHM of primary beam $\sim$1 arcmin) allows us to search for other millimetre-bright galaxies in the vicinity of these quasars. We have detected two more galaxies with molecular line emission - one located 21$\arcsec$ ($\sim$170 kpc at $z=2.5$) to the north of the quasar and the other 11.5$\arcsec$ ($\sim$90 kpc at $z=2.5$) to the south of the quasar. We name these galaxies G1234-N and G1234-S respectively. The northern source G1234-N is also detected in the dust continuum. The CO line profiles and line maps can be seen in Fig. \ref{fig:1234NS} and we list the positions of the two galaxies (as measured from the centroid of the CO emission) together with continuum flux densities and CO line intensities in Table \ref{tab:1234NS}. As the galaxies are not located at the phase center, we determine the scaling that needs to be applied to the flux densities in order to account for the primary beam correction. This scaling is 1.02 for G1234-N and 1.01 for G1234-S and the flux densities in Table \ref{tab:1234NS} have been adjusted accordingly. In light of the new detections of these two galaxies, we also note that the \textit{Herschel} SPIRE fluxes for ULASJ1234+0907 in \citet{Banerji:14} are almost certainly overestimated as all three galaxies fall within the \textit{Herschel} beam at these wavelengths. Reducing the \textit{Herschel} SPIRE fluxes for ULASJ1234+0907 to $\sim$40 per cent of their measured values (to account for the flux density ratio measured between the quasar and the other millimetre bright galaxies with ALMA), means the FIR luminosity of this source decreases by $\sim$0.5 dex. Even accounting for this reduction in the FIR luminosity, ULASJ1234+0907 still has a measured SFR of $\sim$1000 M$_\odot$ yr$^{-1}$. Note, the dust mass for ULASJ1234$+$0907 in Table \ref{tab:derived} is calculated from the ALMA continuum detection only and is therefore not affected by the \textit{Herschel} blending.


\begin{table*}
\begin{center}
\caption{Properties of the two millimetre-bright galaxies detected in the field-of-view of the quasar ULASJ1234+0907 at $z=2.503$}. 
\begin{tabular}{lccccc}
Object & RA & Dec & z$_{\rm{CO}}$ & S$_{\rm{3mm}}$ / $\mu$Jy & I$_{\rm{CO}}$ / Jy km s$^{-1}$ \\
\hline
G1234-N & 12:34:26.36 & 09:08:06.1 & 2.514 & 97$\pm$5 & 0.79$\pm$0.14 \\
G1234-S & 12:34:27.31 & 09:07:43.1 & 2.497 & $<$40 (2$\sigma$) & 0.39$\pm$0.12 \\
\hline
\end{tabular}
\label{tab:1234NS}
\end{center}
\end{table*}

The CO line profile in G1234-N is well fit by a single Gaussian with FWHM=530$\pm$100 km s$^{-1}$, while the line in G1234-S is fit by a Gaussian with FWHM=540$\pm$130 km s$^{-1}$. Based on these CO and dust continuum detections, we can estimate the dust masses and gas masses for the two companion galaxies. For the gas masses, we assume an excitation ratio of $r_{32/10}=0.67$, representative of SMGs, and a CO-to-H$_2$ conversion factor of $\alpha_{\rm{CO}}$=0.8 M$_\odot$ (K km s$^{-1}$ pc$^2$)$^{-1}$. The gas masses for G1234-N and G1234-S are (3.3$\pm$0.6)$\times10^{10}$M$_\odot$ and (1.9$\pm$0.5)$\times10^{10}$M$_\odot$ respectively. Based on the dust continuum detection in G1234-N and assuming a dust temperature of 30K and a dust emissivity index, $\beta=1.5$ (e.g. \citealt{Casey:14}), the dust mass is $\sim$1.4$\times10^9$M$_\odot$ and the SFR is $\sim$600 M$_\odot$ yr$^{-1}$.

The two galaxies G1234-N and G1234-S while clearly not directly interacting with the quasar, are likely to be part of the same overdensity. Several protoclusters have been found at $z>2$ with overdensities of galaxies extending tens of Mpc across e.g. \citep{Venemans:07, Carilli:11, Casey:15, Capak:11}. 

\subsubsection{ULASJ2315+0143: Spatially Resolved Studies of the Molecular Gas}

\label{sec:2315}

As seen in Section \ref{sec:dynamics} and Fig. \ref{fig:specast}, in ULASJ2315+0143 we find evidence that the gas emission is spatially resolved. The deconvolved source size derived from the collapsed channel map summed across the full velocity extent of the line, is (2.84$\times$1.36) $\pm$ (0.30$\times$0.53) arcsec which corresponds to $\sim$23 kpc along the major axis at the quasar redshift. At the current resolution of our data, this size measurement is still uncertain. However in Section \ref{sec:sizes} we show that the large inferred size is robust to how the visibilities are weighted to create the final images. We compare this size estimate to the prototype $z\sim4$ SMG GN20, which has one of the most extended gas reservoirs seen in high redshift galaxies - 14$\pm$4 kpc in diameter as mapped using the CO(2-1) emission \citep{Hodge:12}. Thus, the warm, excited gas in ULASJ2315$+$0143 appears to be even more extended compared to what is seen in lower excitation CO lines in the most extreme SMGs. However, the constraint on the source size in ULASJ2315+0143, is very similar to the size of the beam, and further higher resolution observations are needed to determine whether this extended gas reservoir is associated with a single source. 

\begin{figure}
\begin{center}
\vspace{-3cm}
\begin{tabular}{c}
\vspace{-5.5cm}
\includegraphics[scale=0.4]{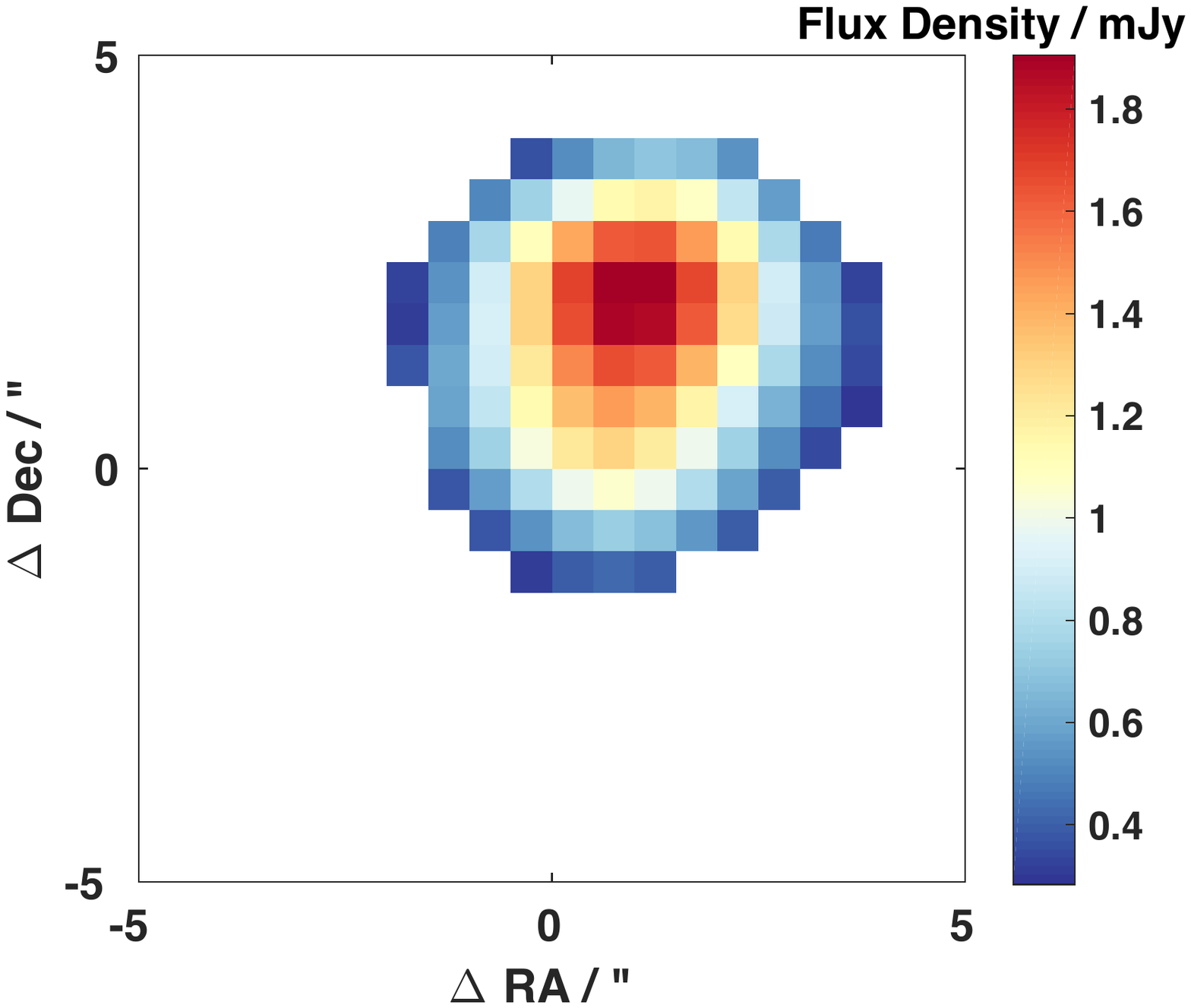} \\
\vspace{-5.5cm}
\includegraphics[scale=0.4]{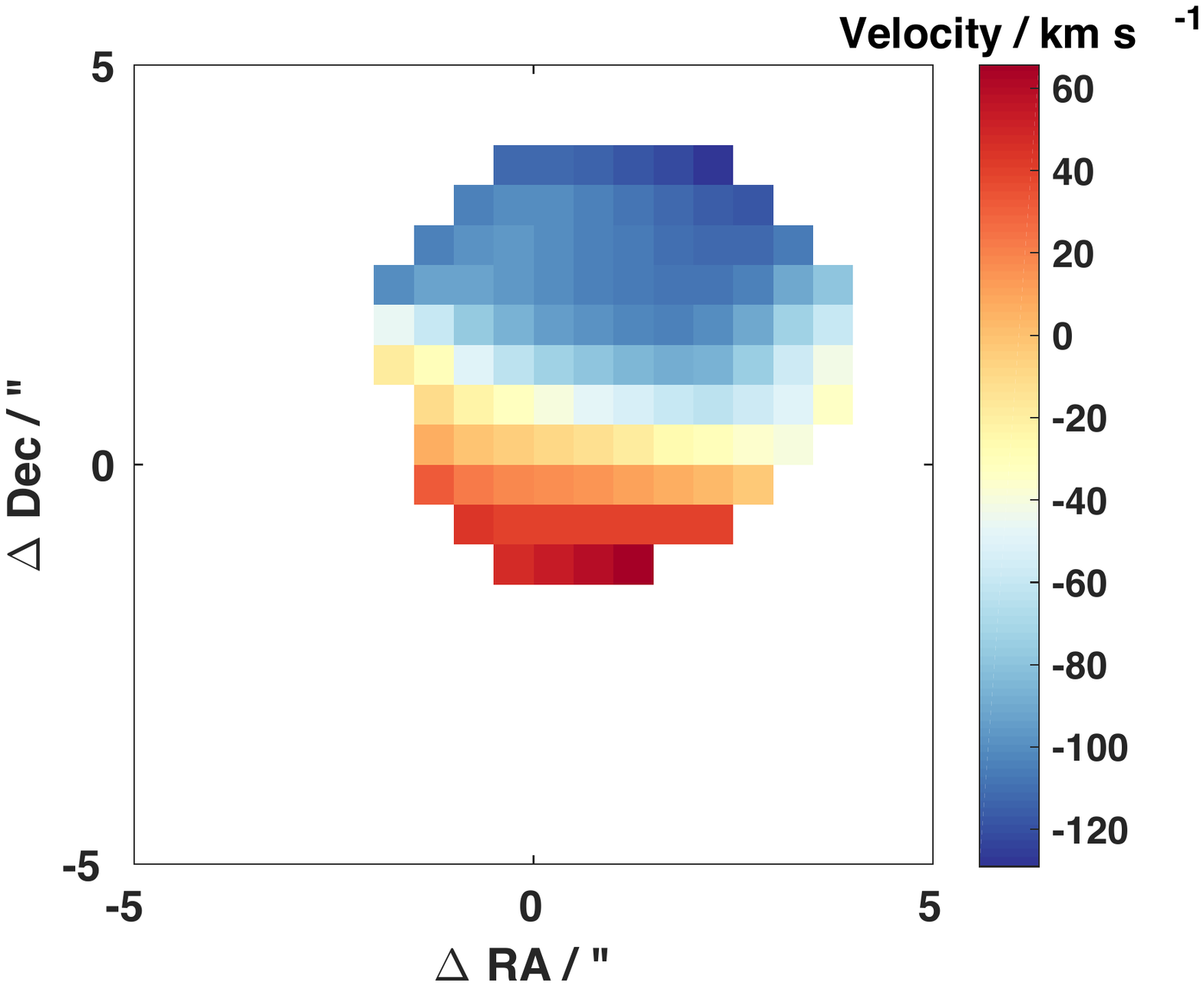} \\
\includegraphics[scale=0.4]{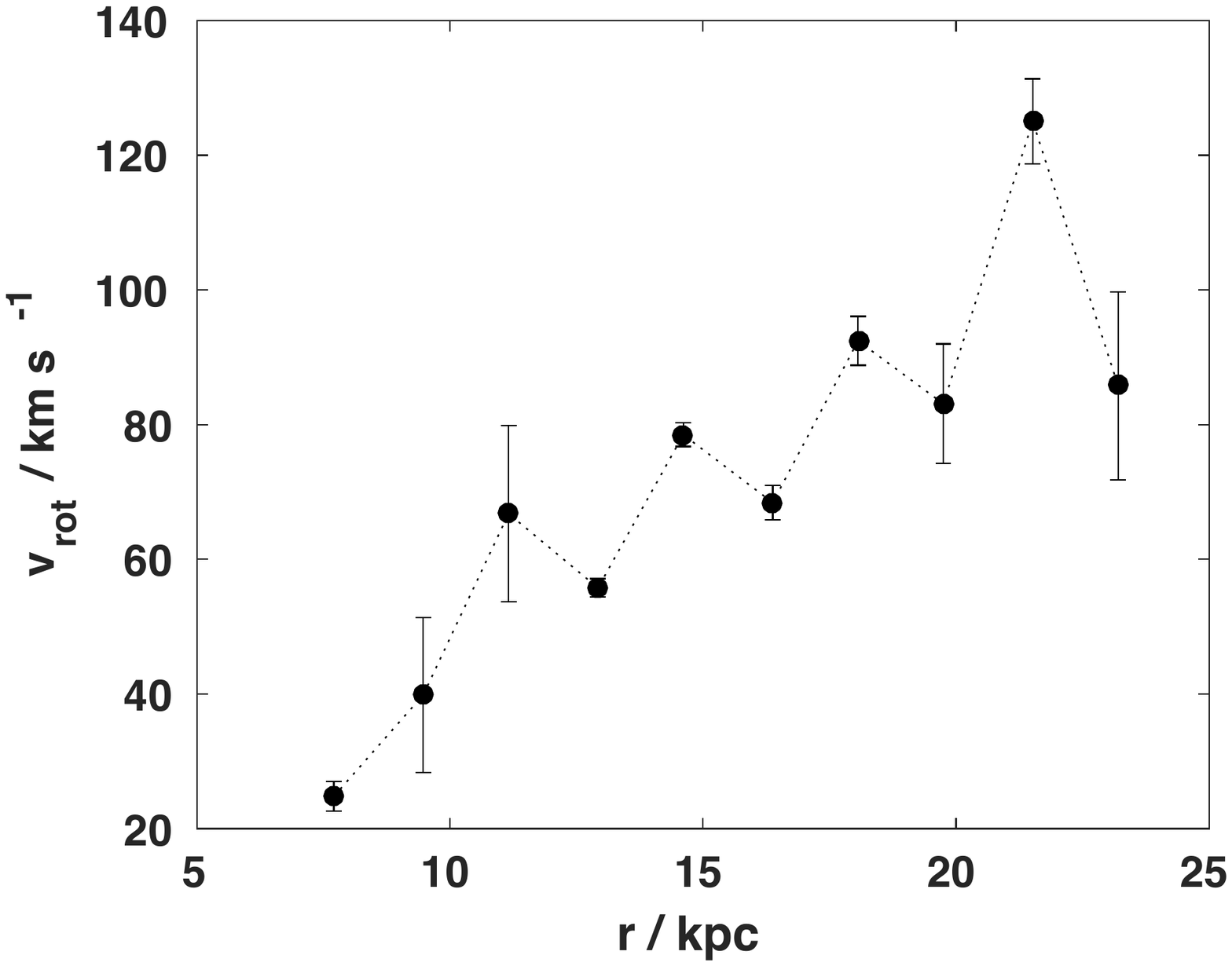} \\
\end{tabular}
\end{center}
\vspace{-3cm}
\caption{Zeroth (top) and first moment (middle) maps of the CO emission in ULASJ2315+0143 derived by fitting a single Gaussian to the line emission in each pixel. The systemic velocity has been subtracted from the velocity field shown in the middle panel, which therefore traces the rotational velocity only. The bottom panel shows the rotation curve derived by fitting a tilted ring gas disk dynamical model to the velocity field shown in the middle panel.}
\label{fig:2315_spatial}
\end{figure}

We construct zeroth- and first-moment maps of the line emission by fitting a single Gaussian profile to the line emission in every pixel, retaining pixels in the central region (i.e. avoiding the edges of the field-of-view) and where the S/N of the line is at least 3. These maps can be seen in Fig. \ref{fig:2315_spatial} and a velocity gradient is evident across the source.  

We fitted a tilted ring disk model to the velocity field shown in Fig. \ref{fig:2315_spatial}. In such an analysis, the gas emission is modelled as series of concentric rings with inclination, $i$, position angle, P.A., systemic velocity $v_{\rm{sys}}$, centre $[x_0,y_0]$ and rotational velocity, $v_{\rm{rot}}$. The line of sight velocity at any position $[x,y]$ can then be described on a ring with radius, $r$ as:

\begin{equation}
v(x,y)=v_{\rm{sys}} + v_{\rm{rot}}(r) sin(i) cos(\theta)
\label{eq:vel}
\end{equation}

\noindent where the angle, $\theta$ is related to P.A. in the plane of the sky by

\begin{equation}
cos(\theta)=\frac{-(x-x_0)sin(P.A.)+(y-y_0)cos(P.A.)}{r}
\end{equation}

\begin{equation}
sin(\theta)=\frac{-(x-x_0)cos(P.A.)-(y-y_0)sin(P.A.)}{rcos(i)}
\label{eq:angle}
\end{equation}

\noindent We used the \textit{rotcur} task within \textsc{gipsy} in iterative mode for the fitting. The expansion velocity was set to zero and the dynamical centre of all the rings was set to be identical. The rings have widths ranging from 0.21-3$\arcsec$. The source has a maximum radial extent of 2.9$\arcsec$ or $\sim$23 kpc. We fixed the P.A. to be that measured from the velocity field by looking at the axis connecting the maximum and minimum velocities. This P.A. is measured to be 352.3$^{\circ}$. The \textit{rotcur} task was run 15 times with one to fifteen rings covering the entire radius with only the systemic velocity and then the inclination angle, $i$, as free parameters. The systemic velocity and $i$ were then determined from the average of all successful rings where a successful ring is defined to be one with errors on $i$ of $<$20$^{\circ}$. We have checked that using an error weighted average instead makes very little difference to the model fit parameters. Finally, we ran the task with $v_{\rm{rot}}$ as the only free parameter and used the results from the most successful rings (defined as those with uncertainties on $v_{\rm{rot}}<3 \times v_{\rm{rot}}$) in a single run to produce a model velocity field and a rotation curve for the source. The best fit rotation curve can be seen in the bottom panel of Fig. \ref{fig:2315_spatial}. The best fit tilted ring model has $i$=57$\pm$8$^{\circ}$. 

Given the resolution of our data, the model fails to constrain the internal dynamics of this source at $r \lesssim 7$ kpc and the rotation curve shown in Fig. \ref{fig:2315_spatial} is noisy and should be considered as illustrative. However, given that these data were taken at the lowest spatial resolution available in ALMA Band 3, the fact that dynamical modelling is even possible with this data highlights the enormous promise of gaining detailed insight into the dynamics of such high redshift quasar host galaxies with higher resolution ALMA observations. We defer a more detailed discussion of the dynamical properties of this source to future papers. 

We can estimate the dynamical mass from the rotation curve shown in Fig. \ref{fig:2315_spatial}. The maximum rotational velocity, $v_{\rm{max}}$=125$\pm$6 km s$^{-1}$ at r=21.5 kpc. Using $M(<r) \rm{sin}^2(i)=v_{\rm{max}}^2 r / G$, we get a dynamical mass of (7.82$\pm$0.02)$\times$10$^{10}$M$_\odot$ which is a factor of $\sim$2 lower than that derived using the spectroastrometry method in Section \ref{sec:dynamics}. Correcting for the full range in inclination angles allowed by the fit, M$_{\rm{dyn}}$=(1.0$-$1.4)$\times$10$^{11}$M$_\odot$.  

\subsubsection{ULASJ0123$+$1525: Distinct Kinematic Components}

ULASJ0123+1525 shows two distinct peaks in its CO line profile, which could indicate the presence of kinematically distinct components in the gas emission from this galaxy. A double Gaussian fit yields a significantly better reduced $\chi^2$ value compared to a single Gaussian (1.2 versus 2.6 for the single Gaussian fit). The two Gaussians have FWHM=240$\pm$80 km s$^{-1}$ and 570$\pm$70 km s$^{-1}$ respectively and are separated in velocity by 260$\pm$60 km s$^{-1}$. At the spatial resolution of our data, there is no evidence for a spatial offset between these velocity components. Similar double peaked line profiles have been observed in SMGs (e.g. \citealt{Greve:05, Tacconi:06, Bothwell:13}) and could arise from a galaxy merger, a rotating gas disk with a large inclination angle relative to the sky plane or large scale inflows and outflows affecting the molecular gas. Further high resolution observations of the molecular gas would help discriminate between these scenarios.   

\section{Discussion}

\label{sec:discussion}

\subsection{Comparison to Unobscured ``Blue" Quasars and X-Ray AGN}

\label{sec:comp_blue}

In order to understand whether our reddened quasars are indeed a distinct population in terms of their molecular gas and dust continuum properties, we now draw direct comparisons to observations of the molecular gas and dust continuum in samples of blue, unobscured quasars with comparable luminosities and redshifts. To date, the largest surveys for cold dust in high luminosity quasars at $z\sim2-3$, are still those conducted using the SCUBA bolometer (at 850$\mu$m) on the James Clerk Maxwell Telescope and using the MAMBO bolometer (at 1200$\mu$m) on the IRAM 30-m telescope (e.g. \citealt{McMahon:94, McMahon:99, Isaak:02, Priddey:03, Omont:03}). The SCUBA surveys have typically probed down to 3$\sigma$ flux limits of $\sim$7 mJy at 850$\mu$m while the MAMBO survey reaches sensitivities of $\sim$2-4 mJy (3$\sigma$). 

In the SCUBA studies, only $\sim$20 per cent of the optically unobscured, blue quasars are ``submillimetre loud" with 850$\mu$m flux densities of $\gtrsim$7 mJy. We can estimate the 850$\mu$m fluxes for the reddened quasars using the assumptions regarding the dust SED set out in Section \ref{sec:dust}. These 850$\mu$m flux densities are compared to other samples from the literature in Fig. \ref{fig:z_850} and the errorbars reflect the difference in 850$\mu$m flux between assuming T$_{\rm{d}}$=47K and $\beta$=1.6 \citep{Beelen:06, Wang:08} and T$_{\rm{d}}$=41K and $\beta$=1.95 \citep{Priddey:01}. Three of the four reddened quasars lie near or below the 2$\sigma$ limit for the SCUBA surveys. From this initial sample of reddened quasars, and with the assumptions regarding their dust SED set out above, $\sim$25 per cent of the reddened quasars appear to be ``submillimetre loud", similar to the submillimetre loud fraction of blue quasars. We can also compare the submillimetre and millimetre flux densities for our reddened quasars with the stacked results from the SCUBA surveys, which suggest that the average 850$\mu$m flux density of a flux limited sample of blue quasars is 1.9$\pm$0.4 mJy at $z\sim2$ \citep{Priddey:03} and 2.0$\pm$0.6 mJy at $z\sim4$ \citep{Isaak:02}. Based on our ALMA detections, our reddened quasars would all have 850$\mu$m fluxes that are higher than these average properties for the blue quasars as can be seen in Fig. \ref{fig:z_850}. The larger 850$\mu$m fluxes could suggest that luminous reddened quasars have higher dust luminosities and are therefore, on average, more likely to be hosted in starburst galaxies relative to their unobscured blue counterparts. However, larger samples of reddened quasars with cold dust continuum measurements and better constraints on their dust SED are clearly required before any robust conclusions can be drawn.

We also compare the submillimetre fluxes of our reddened quasars to those of X-ray selected AGN. X-ray surveys of high redshift AGN have generally covered smaller areas on the sky compared to optical surveys and therefore  probe quasars and AGN at lower luminosities. Considering only the most X-ray luminous, high redshift quasars in the COSMOS field with SCUBA-2 850$\mu$m observations from the SCUBA-2 Cosmology Legacy Survey, the stacked 850$\mu$m fluxes are 0.7$\pm$0.2 mJy for Type 1 AGN and 1.3$\pm$0.3 mJy for Type 2 AGN \citep{Banerji:15b}, once again lower than those measured for the reddened quasars in this work. There are however a number of X-ray AGN that are individually detected at 850$\mu$m in \citet{Banerji:15b} with 850$\mu$m flux densities of $\sim$4-10 mJy, comparable with the inferred submillimetre fluxes of our reddened quasars. Previous SCUBA surveys of X-ray absorbed and X-ray unabsorbed quasars have suggested a link between cold dust emission and X-ray obscuration \citep{Page:01, Stevens:05} and we note that for those reddened quasars where we have X-ray spectra (e.g. \citealt{Banerji:14}), moderate column densities of $\sim$10$^{22}$ cm$^{-2}$ are present.    

Surveys for molecular gas in blue quasars have focussed on the subset of the population that are (sub)millimetre loud, in contrast to our ``blind" search for CO in the reddened quasar population. So far six of our near infrared selected, heavily reddened quasars from the B12 and B15 samples have been observed in CO - four in this paper and one each in \citet{Feruglio:14} and \citet{Brusa:15}. Despite not being selected a-priori to be FIR/mm bright, all six quasars have been detected in CO. As stated earlier, most optical quasars that have been observed in CO were known a-priori to be FIR luminous therefore increasing the probability of a CO detection. \citet{Coppin:08} have looked for CO in a sample of 10 quasars of comparable redshift to our reddened quasar sample. Three quasars in their sample were detected in blank field submm surveys over much smaller areas of sky than the wide-field near infrared surveys that have been used to select our reddened quasars (B12,B15), and are therefore lower luminosity. The remaining seven quasars in the \citet{Coppin:08} sample, are brighter than the $K$-band flux limit imposed on our reddened quasar searches in B12 and B15 and can therefore be directly compared to our sample. From these seven FIR bright optical quasars, five have been detected in CO with the limits on the CO line luminosity for the remaining two placing them below the CO line luminosities of all six of our reddened quasars. Presumably, starting from a complete sample of unobscured quasars (i.e. not just those known a-priori to be millimetre bright) would result in a smaller fraction being detected in CO. The black hole masses of our reddened quasars and the UV-luminous \citet{Coppin:08} quasars are comparable given the typical uncertainties in these black hole mass estimates. Hence, the CO-detection fraction in our sample suggests that a larger fraction of luminous reddened quasars have significant molecular gas reservoirs compared to UV luminous blue quasars of comparable luminosity and mass.

In B15 we demonstrated that our reddened quasars outnumber blue quasars at the highest luminosities whereas the reddened quasar population is subdominant as we approach more typical luminosities around L$^*$. One interpretation of these trends put forward in B15 is that the most luminous, massive black holes have a longer duty cycle associated with their assembly or growth phase and therefore are seen as dust obscured for a longer period of time. As all our ALMA targets are drawn from the high luminosity, massive end of the reddened quasar population, the detection of significant amounts of cold dust and molecular gas in these systems fits in with this overall picture. In such an evolutionary sequence, (sub)millimetre-loud blue quasars are presumably a later, more evolved phase when significant gas reservoirs and star formation are still detectable but some of the dust has been cleared out. The (sub)millimetre-faint blue quasars may then correspond to an even more advanced phase when significant star formation has ceased either because the gas supply has been depleted, or as a result of AGN feedback. 

\subsection{Comparison to Mid-Infrared Luminous AGN}

\label{sec:comp_mir}

\begin{figure}
\begin{center}
\centering
\vspace{-3.7cm}
\hspace{-1.0cm}
\includegraphics[scale=0.45]{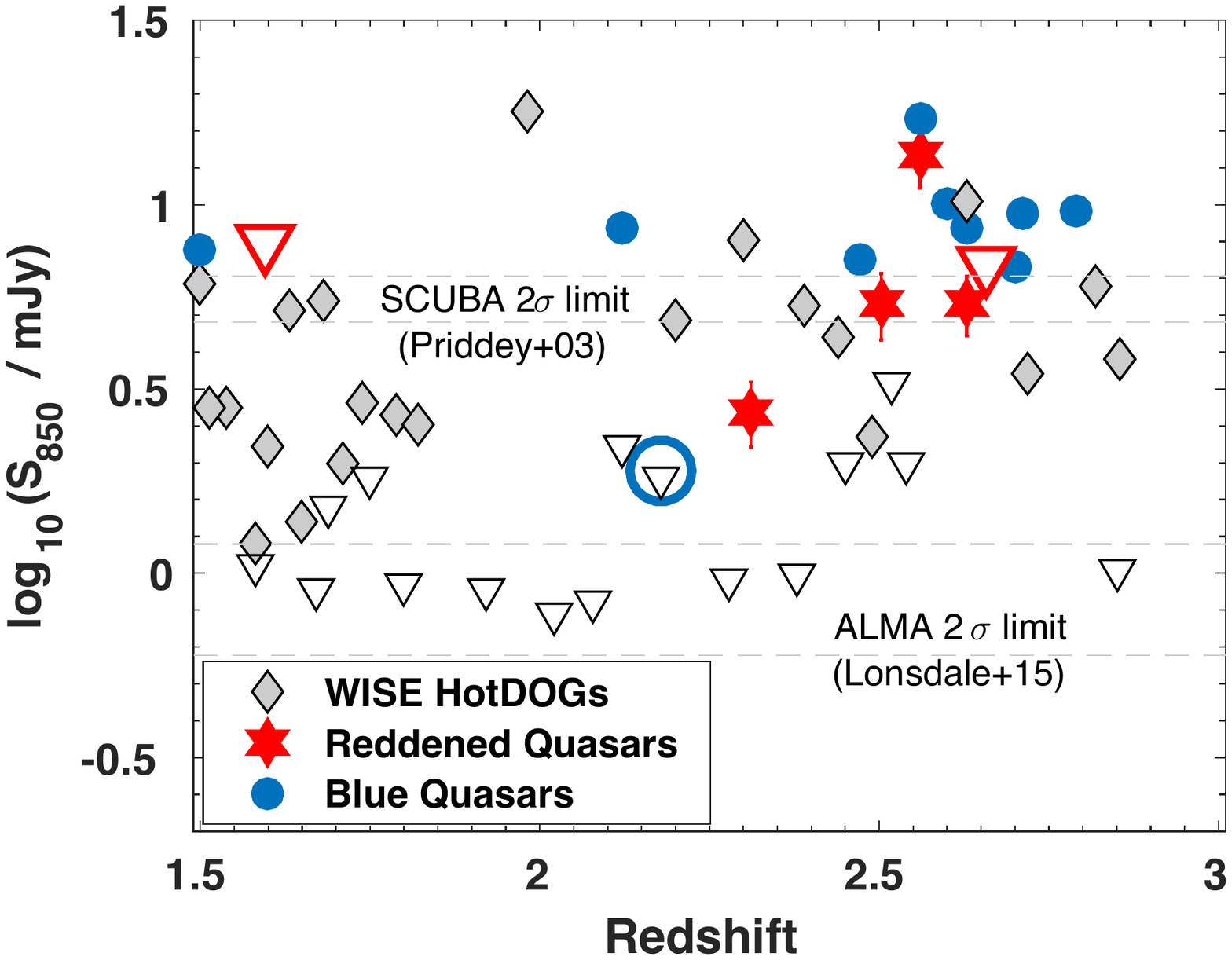} \\
\vspace{-3.5cm}
\caption{Redshift versus 850$\mu$m flux density for blue quasars \citep{Priddey:03}, reddened quasars (This Work, \citealt{Feruglio:14, Brusa:15}) and \textit{WISE} HotDOGs \citep{Lonsdale:15, Jones:14, Jones:15}. In all cases downward triangles denote upper limits and the large blue open circle shows the stacked 850$\mu$m flux of the blue quasars from \citet{Priddey:03}. The dashed lines mark the range in 2$\sigma$ flux limits for the SCUBA survey \citep{Priddey:03} and the ALMA survey \citep{Lonsdale:15}.}
\label{fig:z_850}
\end{center}
\end{figure}

Under the premise that dust reddened quasars are a phase in massive galaxy evolution preceding the unobscured quasar phase, and that this phase is associated with high levels of star formation and high rates of accretion onto the supermassive black hole, we would predict that populations of high luminosity AGN that are even more obscured than our reddened quasars may have similar or higher levels of star formation in their host galaxies. These most highly obscured AGN might therefore be expected to be characterised by high 850$\mu$m fluxes. The most suitable sample of mid-infrared luminous, heavily obscured AGN with comparable AGN luminosities and redshifts as our reddened quasars are the so called AGN dominated Hot Dust Obscured Galaxies or HotDOGs that have emerged from the \textit{WISE} All Sky Survey (e.g. \citealt{Eisenhardt:12}). We use as the basis for comparison, the SCUBA-2 850$\mu$m and ALMA 870$\mu$m flux densities of the spectroscopically confirmed \textit{WISE} HotDOGs at $1.5 < z < 3.0$ from \citet{Jones:14, Jones:15} and \citet{Lonsdale:15}. These studies provide a sample of 42 HotDOGs with secure spectroscopic redshifts overlapping the redshift range of both our reddened quasar sample and the \citet{Priddey:03} optical quasars. Once again we extrapolate our reddened quasar 3mm flux densities using the SED assumptions above to predict their 850$\mu$m fluxes. A comparison of the 850$\mu$m flux densities can be seen in Fig. \ref{fig:z_850}.

Out of the 42 WISE HotDOGs, 19 are detected at high significance at 850/870$\mu$m - a detection fraction of $\sim$45 per cent. If we consider the fraction detected to the SCUBA limits of $\sim$7 mJy, the number drops to only 3 - i.e. just 7 per cent of the population. As seen in Fig. \ref{fig:z_850}, in many cases the limits on the submillimetre fluxes obtained for the \textit{WISE} HotDOGs, place them well below the flux densities of our reddened quasars. Thus, as concluded in \citet{Lonsdale:15} and several other recent \textit{WISE} papers, these HotDOGs on average do not show evidence for the dust luminosities being dominated by cold dust heated by star formation and instead, it is likely that most of the dust heating in these galaxies comes from the AGN. The \textit{WISE} HotDOGs therefore do not seem to fit our simple evolutionary picture linking obscured and unobscured quasars. As suggested in \citet{Eisenhardt:12}, these \textit{WISE} AGN could perhaps represent a class of objects where the peak of AGN activity precedes that of star formation. Alternatively, the higher line-of-sight extinction seen in these sources could simply reflect a difference in viewing angle between these \textit{WISE} AGN and both the reddened and unobscured broad line quasar populations. 

There have been relatively few studies of CO molecular gas in the \textit{WISE} population so far but \citet{Wu:14} have looked for $^{12}$CO(3-2) and $^{12}$CO(4-3) in two of these galaxies. Neither source is detected in their observations but the limits on the gas mass from that study are consistent with the low end of the gas masses we have derived for the reddened quasar population in this paper. \\

It is clear from Fig. \ref{fig:z_850} that the number of dust continuum observations for optical and \textit{WISE}-selected AGN in much larger than the number available for reddened quasars. With a large spectroscopic sample of the reddened quasars now in place, it is imperative that we extend the investigations carried out in this work to larger samples in order to gain a coherent picture of the cold dust and molecular gas emission and star formation in these different quasar populations covering a very wide range in line-of-sight extinction.

\subsection{Star Formation Efficiencies and Gas Depletion Timescales}

\begin{figure}
\begin{center}
\vspace{-3.9cm}
\hspace{-2.1cm}
\includegraphics[scale=0.5]{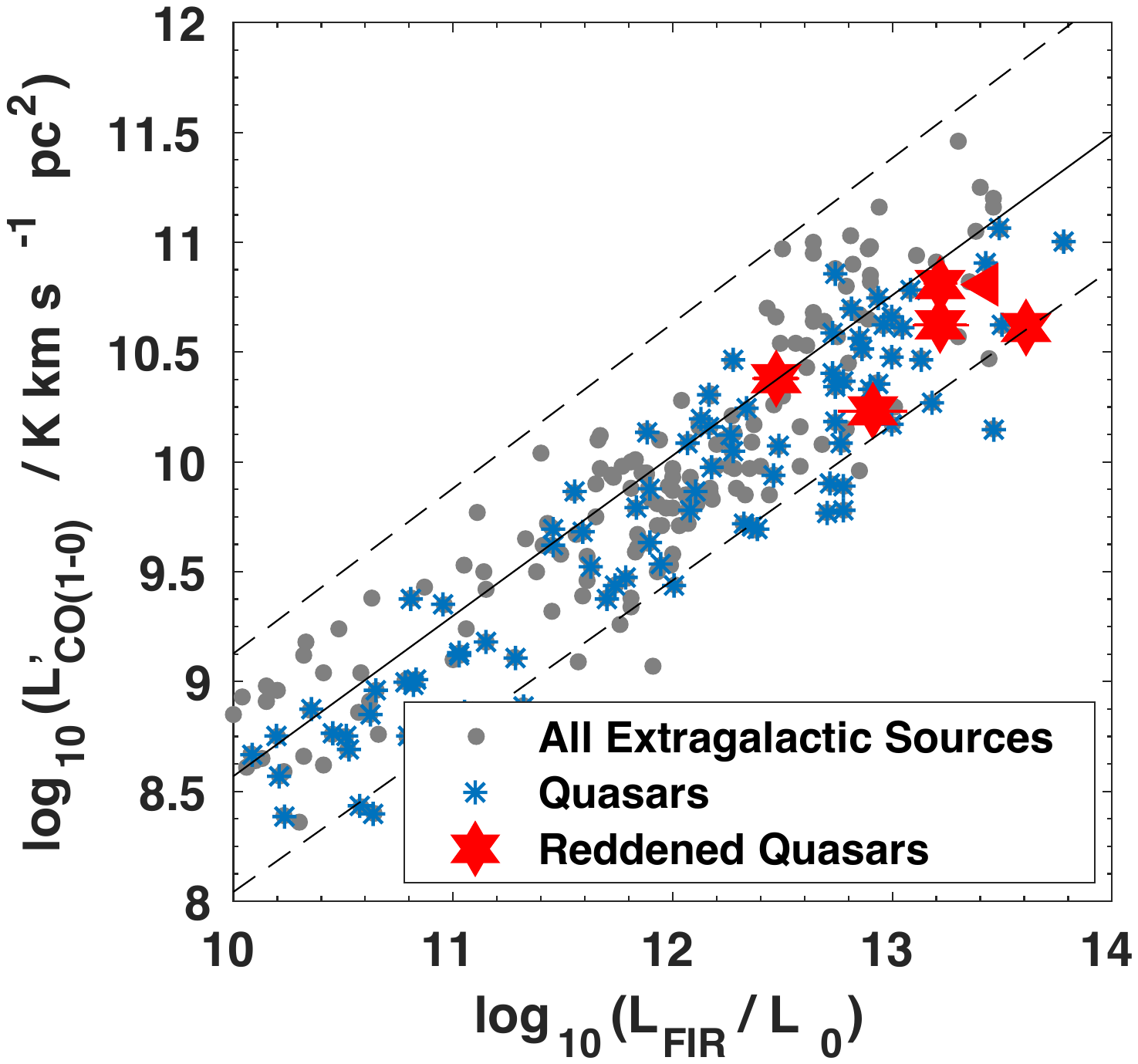}
\vspace{-4cm}
\caption{FIR luminosity versus CO line luminosity for extragalactic sources from the literature including quasars \citep{Heywood:13}, compared to our sample of reddened quasars presented in this paper and in \citet{Brusa:15} and \citet{Feruglio:14}. The infrared luminosities for the reddened quasars have been calculated as discussed in Section \ref{sec:dust}. The lines show the best-fit relation from \citet{Carilli:13} given in Eq. \ref{eq:LIR_LCO} together with the dispersion in this relation.}
\label{fig:LIR_LCO}
\end{center}
\end{figure}

Having detected both cold dust and molecular gas in dust reddened quasars, we can attempt to compare the star formation efficiencies and gas depletion timescales to those measured in high redshift starburst galaxies and quasars. We use as the basis for our comparisons, the compilations of CO detections in high redshift galaxies and quasars published by \citet{Carilli:13} and \citet{Heywood:13}. While these compilations are not necessarily exhaustive and complete, they can be taken to be representative of current observational samples. Throughout this analysis we work with the full sample of six heavily reddened quasars with CO detections - i.e. the four in this paper plus the two from \citet{Feruglio:14} and \citet{Brusa:15}. For the last two quasars we have re-derived the CO line luminosities using the same assumptions regarding the CO excitation as in this work (see Table \ref{tab:derived}). In Fig. \ref{fig:LIR_LCO} we show the CO line luminosity, L$^{'}_{\rm{CO(1-0)}}$, versus the infrared luminosity for all extragalactic sources from \citet{Heywood:13} and the six heavily reddened quasars in Table \ref{tab:derived}.  Also shown is the best-fit relation from \citet{Carilli:13}:

\begin{equation}
\rm{log(L_{IR})}=1.37(\pm0.04)\times\rm{log(L^{'}_{CO}}-1.74(\pm0.40))
\label{eq:LIR_LCO}
\end{equation}

\noindent The reddened quasars in general have higher infrared luminosities or star formation rates for a given gas mass, than predicted by this relation, which has been derived from the entire distribution of high redshift galaxies in the literature. This is consistent with a high star formation efficiency in these reddened quasar hosts compared to other extragalactic sources The infrared luminosities are however broadly consistent with those of other high redshift quasar host galaxies and the scatter in the observed relation shown in Fig. \ref{fig:LIR_LCO} is large. We caution also that the infrared luminosities in the literature have been calculated using heterogenous methods both in terms of the wavelength range used to compute these luminosities and the assumptions made regarding the form of the dust SED. 

To gain further insight into the similarities and differences in the star formation efficiencies between SMGs, quasars and reddened quasars, we now compare both the FIR luminosity distributions and CO luminosity distributions over the same redshift range, using a KS-test to determine the probability that they are drawn from the same parent population. There is no evidence for a difference in the FIR luminosity distributions of the three samples when the same assumptions are made regarding the dust temperature and emissivity index of the FIR SED. Comparing the L$^{'}_{\rm{CO}}$ luminosities for the three samples however, we find that quasars and SMGs have a 3 per cent probability of being drawn from the same distribution, highlighting that the molecular gas reservoirs may indeed be different in the two samples. Reddened quasars on the other hand have a 32 per cent probability of having CO luminosities drawn from the same luminosity distribution as SMGs and a 54 per cent probability of being drawn from the same luminosity distribution as optical quasars so their gas properties are consistent with both SMGs and optical quasars. 

We calculate typical gas depletion timescales using the gas masses and star formation rates in Table \ref{tab:derived}. Regardless of whether the CO based or dust based gas masses are used, the gas depletion timescales are very short - $\sim5-20$ Myr. The black hole mass accretion rates of $\sim$100-500 M$_\odot$yr$^{-1}$ (B12,B15) imply black hole gas consumption timescales that are at least an order of magnitude longer - $\sim$100-1000 Myr. The black holes are already very massive - $\sim10^9-10^{10}$M$_\odot$ - suggesting we are witnessing a phase when the galaxies are running
out of gas (to further fuel the black hole) following the main period of black-hole mass assembly. The reddened quasars could thus be plausible progenitors of the largest supermassive black holes seen in the Universe today, which have measured black hole masses of $>$10$^{10}$M$_\odot$ (e.g. NGC 3842 and NGC 4899; \citealt{McConnell:11}). 


\subsection{$^{12}$CO(3-2) Sizes}

\label{sec:sizes}

A detailed discussion of the sizes of the CO emitting regions in reddened Type 1 quasars is beyond the scope of this paper and will have to wait for higher resolution observations of the gas distribution in these sources. However, we have seen that the CO emission is already resolved in one of our reddened quasars - ULASJ2315$+$0143 (Section \ref{sec:2315}) - with an implied source diameter of $>$20 kpc. There are hints also that the gas emission is marginally resolved in ULASJ1234+0907 and VHSJ2101$-$5943. We used \textsc{casa} to fit a 2-D Gaussian (in RA, Dec) to the CO maps constructed by summing all the channels covering the full extent of the line. The deconvolved source sizes along the major axis are (2.53$\pm$0.52) arcsec $\equiv$(20$\pm$4 kpc) and (1.34$\pm$0.29) arcsec $\equiv$(11$\pm$2) kpc for ULASJ1234+0907 and VHSJ2101$-$5943 respectively. We note that the images used for the size estimates were generated using \textit{natural} weighting, which results in strong sidelobes in the image plane. We have therefore checked that the sizes quoted here are consistent within the error bars with the deconvolved sizes estimated from images generated using a \textit{Briggs} weighting (robust=0.5), where the effects of the sidelobes should be minimised. Finally, for our most extended and highest S/N source, ULASJ2315$+$0143, we have also checked that the major axis derived from directly fitting to the visibility data using the \textsc{casa} task \textit{uvmodelfit}, is once again consistent with the deconvolved sizes from the image plane quoted here.

The large implied sizes for the warm gas reservoirs in these quasars, while still highly uncertain, are nevertheless at odds with previous observations of gas in quasar host galaxies. In the few cases where the gas emission has been resolved in high redshift quasar host galaxies, it is distributed in compact regions of $<$ few kpc in size (e.g. \citealt{Carilli:02, Walter:04, Riechers:11b, Wang:13, Willott:13}). The conventional interpretation is that the molecular gas in quasar hosts resides in a nuclear region close to the accreting black hole, and its excitation properties can therefore be explained by a single temperature and density gas component. SMGs on the other hand often show evidence for substantially larger reservoirs of low excitation gas $\gtrsim$10 kpc in size (e.g. \citealt{Ivison:11, Riechers:11a, Hodge:12}). Although the excitation properties of the gas in reddened quasars remains to be determined, we tentatively conclude that the CO sizes appear to more closely resemble SMGs rather than high redshift quasar hosts. 

\subsection{Molecular Gas Fractions}

\begin{figure}
\begin{center}
\vspace{-2.2cm}
\includegraphics[scale=0.4]{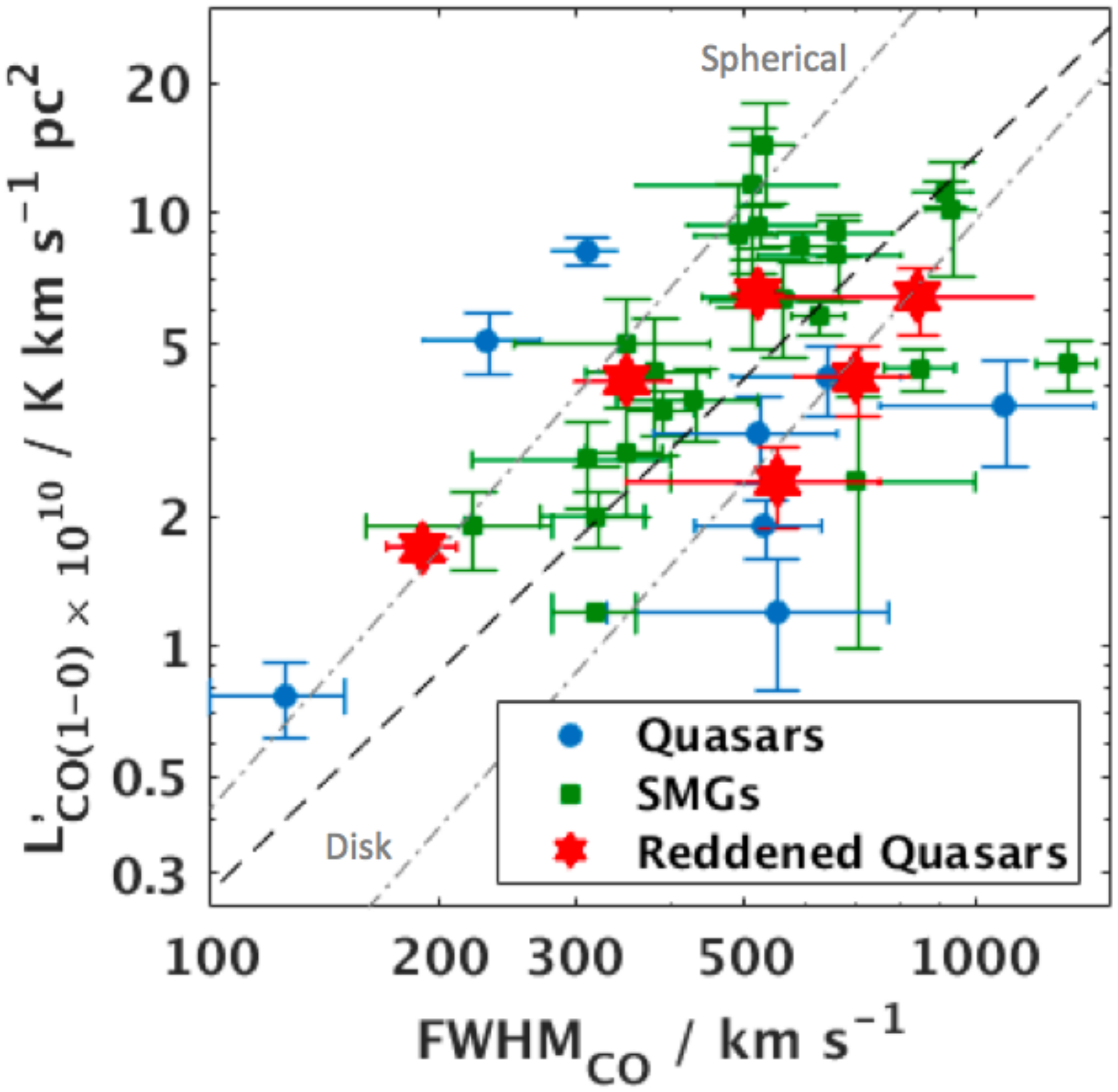}
\vspace{-3cm}
\caption{FWHM of the CO line (proxy for dynamical mass) versus CO line luminosity (proxy for gas mass) for quasars and SMGs from the literature \citep{Bothwell:13, Coppin:08, Simpson:12} compared to our reddened quasars (This work, \citealt{Feruglio:14, Brusa:15}). All sources shown here are unlensed and therefore no magnification correction needs to be applied to these data. The dashed line shows the best-fit relation to unlensed galaxies from \citet{Harris:12} while the dot-dashed lines show the simple relation given by Eq. \ref{eq:LFWHM} for both disk like and spherical geometries.}
\label{fig:FWHM_LCO}
\end{center}
\end{figure}

With the CO detections providing us with molecular gas masses and the CO velocity widths providing a crude estimate of the dynamical masses, we can estimate the molecular gas fractions and compare directly to gas fractions in both SMGs and optical quasars. A KS-test comparing the observed FWHM of the CO lines in SMGs, optical quasars and reddened quasars reveals no evidence for a difference in the distributions between the three populations. If the three populations indeed form an evolutionary sequence with the optical quasars representing more evolved systems and SMGs representing the earlier gas rich stages of massive galaxy formation, we would expect the reddened quasars to have gas fractions that are intermediate between SMGs and optical quasars. In Fig. \ref{fig:FWHM_LCO} we show the FWHM of the CO line - a proxy for the dynamical mass - versus the CO(1-0) luminosities - a proxy for gas mass. We compare our sample to optical quasars from \citet{Coppin:08} and \citet{Simpson:12} and to the SMGs from \citet{Bothwell:13}, selecting quasars and SMGs that are unlensed only. We also show the scaling relation between these two quantities derived for unlensed sources by \citet{Harris:12}, as well as a simple relation that equates the gas mass to the dynamical mass as follows:

\begin{equation}
L^{'}_{\rm{CO(1-0)}}=\frac{C \sigma^2 R}{G \alpha_{\rm{CO}}}
\label{eq:LFWHM}
\end{equation}

\noindent  where the constant, $C$ depends on the geometry and inclination of the source, $\sigma$ represents the line dispersion $\sim$FWHM/2.35 for a Gaussian line, $R$ is the source size and $G$ is the gravitational constant. We consider two separate cases: (i) $C=2.1$, $R=5$ kpc, $\alpha_{\rm{CO}}$=4.6 M$_\odot$ (K km s$^{-1}$ pc$^2$)$^{-1}$, which is appropriate for disk galaxies and (ii) $C=5$, $R=2$ kpc, $\alpha_{\rm{CO}}$=1 M$_\odot$ (K km s$^{-1}$ pc$^2$)$^{-1}$, representing a virialised, spherical source.

We find that the reddened quasars occupy a region roughly between the majority of the optical quasars and the SMGs in Fig. \ref{fig:FWHM_LCO}. However, the observed FWHM is also sensitive to inclination with thin rotating disks viewed face-on having narrower FWHM. \citet{Bothwell:13} have argued based on the fact that SMGs show a correlation between CO linewidth and line luminosity, that they are primarily thick disks/turbulent ellipsoids where the FWHM is not as dependent on inclination. However, optical quasars, where broad emission lines can be seen in the quasar spectra, may have an inclination bias relative to SMGs although the opening angles needed for detection of broad lines in quasars likely span a large range. If the molecular gas sits in a thin disk in these quasars, the optical quasars with very narrow FWHM that sit above the various lines in Fig. \ref{fig:FWHM_LCO}, could be seen more face-on. On the other hand, if the gas morphologies are more turbulent and inhomogenous - e.g. as would be expected for merging systems - the FWHM is unlikely to be as dependent on inclination. Further observations are required to established whether the reddened
quasars have any inclination bias relative to optical quasars and SMGs. We note that there is a general trend among our objects for the more highly obscured quasars to have larger observed CO FWHM - e.g. ULASJ1234+0907, the dustiest quasar in our sample (A$_{\rm{V}} \sim 6$ mag; B12) has the broadest CO line whereas VHSJ2101$-$5943, the least dusty quasar observed (A$_{\rm{V}} \sim 2.5$ mag; B15), has the narrowest CO line, suggesting that some of the reddening in these quasars could be linked to orientation.  However, in Section \ref{sec:2315} we have seen that in at least one of our quasars the gas appears to be distributed in a disk at an inclination angle of 57$^{\circ}$. This is consistent with the inclination angles inferred for SMGs. If SMGs, reddened quasars and optical quasars have similar source
sizes and inclinations on average, for a fixed dynamical mass or CO
FWHM, reddened quasars have gas masses that are intermediate between SMGs and optical quasars, and are therefore on their way to exhausting their gas supply and evolving into optical quasars. Further spatially resolved observations of the reddened quasars to get rotational dynamical parameters as well as internal velocity dispersions, is clearly the next step forward in order to get robust gas fractions in these sources. 

\section{Conclusions}

We have presented ALMA Band 3 observations of a sample of four heavily reddened (A$_{\rm{V}}\sim2.5-6$ mag), high luminosity broad line quasars at $z\sim2.5$ selected from wide field near infrared surveys. The aim of these observations is to test whether such reddened quasars could represent the evolutionary link between starburst galaxies and unobscured, optically bright quasars at high redshifts. All four quasars are detected in both the dust continuum (at observed frame wavelengths of $\sim$3mm) and in the $^{12}$CO(3-2) line. We infer star formation rates, dust and gas masses from these observations and attempt to place these measurements in context with corresponding values for other high redshift galaxy populations:
\begin{itemize}

\item{The dust masses range from 2.5$\times$10$^{8}$M$_\odot$ to 1.2$\times$10$^9$M$_\odot$. The FIR luminosities are $\gtrsim$10$^{13}$L$_\odot$ and the implied star formation rates in the quasar host galaxies is $\sim1200-6000$ M$_\odot$ yr$^{-1}$, suggesting  that these reddened quasars reside in prodigiously star forming host galaxies with very high dust luminosities.}

\item{We estimate gas masses from the measured CO(3-2) line luminosities, assuming an excitation ratio $r_{32/10}$=0.8 and a CO-to-H$_2$ conversion factor of $\alpha_{\rm{CO}}=0.8$ (K km s$^{-1}$ pc$^2$)$^{-1}$. These gas masses are in the range $\sim1-5\times$10$^{10}$M$_\odot$ and the gas-to-dust ratios are in the range $\sim$30-110.}

\item{Instead, adopting a gas-to-dust ratio of 91, as found in nearby
galaxies, dust- and CO-based gas masses agree within a factor of $\sim$2
for three of the quasars. For ULASJ2315+0143, the dust-based
gas mass is $\sim$3 times larger.  The CO- and dust-based masses
could be made consistent for ULASJ2315+0143 if a CO-to-H$_2$ conversion factor of $\alpha_{\rm{CO}}=2.6$ (K km s$^{-1}$ pc$^2$) applies.  Such a value is
intermediate between that found in nearby galaxies/high-redshift
star-forming disks and in nuclear starbursts/quasar host galaxies.}

\item{The CO linewidths span $\sim200-700$ kms$^{-1}$, comparable to linewidths in
other high-redshift SMGs and quasars. Using several different
estimators, dynamical masses of M$_{\rm{dyn}}$sin$^2$(i) $\sim$3$\times$10$^{10}$M$_\odot$ to $\sim$3$\times$10$^{11}$M$_\odot$ are found.}

\item{ULASJ2315+0143 ($z = 2.561$) shows evidence for spatially resolved CO
emission over $\sim$20 kpc, with a strong velocity gradient across the source. We demonstrate that the velocity field is consistent with the presence of a large, rotating gas disk although other scenarios such as mergers cannot be ruled out at the resolution of the current data. ULASJ2315$+$0143 is the quasar with the largest discrepancy between its dust and CO based gas masses, and where there is evidence for $\alpha_{\rm{CO}}>2$ (K km s$^{-1}$ pc$^2$)$^{-1}$.}

\item{The sizes of the $^{12}$CO(3-2) emitting regions appear to be large ($\gtrsim$ 10 kpc) hinting at the presence of extended reservoirs of warm gas. Gas emission in high-redshift, unobscured quasars is generally
compact ($<$few kpc). If extended emission is confirmed via higher
resolution observations, the reddened quasars would instead be more
similar to the properties of SMGs.}

\item{ULASJ1234+0907 ($z=2.503$) has two other millimetre bright galaxies at the same redshift in the field-of-view, with implied gas masses of a few times 10$^{10}$M$_\odot$. The two galaxies are located $\sim$90 kpc and $\sim$170 kpc from the quasar. ULASJ1234+0907 may reside in a significant over-density at $z\sim2.5$.} 

\item{A similar fraction of reddened quasars are ``submillimetre loud" compared to unobscured UV luminous quasars. However the average dust luminosities of reddened quasars would appear to be higher than for the UV luminous quasar population at similar redshifts. Reddened quasars also appear to have higher cold dust luminosities compared to the recently discovered population of Hot Dust Obscured Galaxies from \textit{WISE}. So far, 100 per cent of our reddened quasars (6/6) have been observed to have very large molecular gas reservoirs. A direct comparison to the unobscured quasar population is not possible as surveys of molecular gas in UV luminous quasars have focussed on the subset of these that were already known to be millimetre bright, therefore increasing the probability of detecting molecular gas. Nevertheless, several UV luminous, millimetre bright quasars with comparable black hole masses and AGN luminosities to our sample, have limits on their molecular gas masses that place them below the gas masses seen so far in the reddened quasar population, indicating that these unobscured quasars could indeed correspond to a later evolutionary phase. Further observations of molecular gas in larger samples of both reddened and unobscured quasars matched in luminosity and black hole mass would help confirm this trend.}

\item{Using the CO linewidths in our sample as a proxy for the dynamical mass, and the CO line luminosities as a proxy for gas mass, we find that the reddened quasars have molecular gas fractions that are intermediate between SMGs and optical quasars assuming that all three populations have similar source sizes and inclinations. It therefore seems plausible that reddened quasars are being seen as they are transitioning to optical quasars from starburst galaxies. Their supermassive black holes already appear to be fully assembled and, given the rate of ongoing star formation, they will very quickly exhaust their gas reservoirs and presumably blow out the surrounding gas and dust on their way to evolving into some of the most luminous, unobscured quasars - progenitors of the largest supermassive black holes seen in the Universe today.}

\end{itemize}

\noindent Overall, our observations demonstrate the high scientific return possible from ALMA observations of luminous quasars at the main epoch of galaxy formation. Studies of the host galaxy properties of intrinsically luminous, unlensed quasars at $z\sim2-3$ are still scarce and further observations with ALMA both in terms of assembling larger samples of such sources and obtaining higher spatial resolution detections, will help shed light on the exact nature of the host galaxies of the largest supermassive black holes, seen as they are being assembled in the high redshift Universe.   

\section*{Acknowledgements}

The authors thank the referee for a useful and detailed report which helped improve the paper. MB would like to thank Matt Bothwell and Stefano Carniani for useful discussions. MB acknowledges funding from the UK Science and Technology Facilities Council (STFC) via an Ernest Rutherford Fellowship. GJ is grateful for support from NRAO through the Grote Reber Doctoral Fellowship Program. RGM and PCH acknowledge funding from STFC via the Institute of Astronomy, Cambridge Consolidated Grant. SA-Z acknowledges support from Peterhouse, Cambridge. 

This paper makes use of the following ALMA data: ADS/JAO.ALMA\#2015.1.01247.S. ALMA is a partnership of ESO (representing its member states), NSF (USA) and NINS (Japan), together with NRC (Canada), NSC and ASIAA (Taiwan), and KASI (Republic of Korea), in cooperation with the Republic of Chile. The Joint ALMA Observatory is operated by ESO, AUI/NRAO and NAOJ.




\bibliography{}





\bsp	
\label{lastpage}
\end{document}